\documentclass[fleqn,usenatbib]{mnras}

\usepackage{newtxtext,newtxmath}

\usepackage[T1]{fontenc}

\DeclareRobustCommand{\VAN}[3]{#2}
\let\VANthebibliography\thebibliography
\def\thebibliography{\DeclareRobustCommand{\VAN}[3]{##3}\VANthebibliography}

\DeclareRobustCommand{\DA}[3]{#2}
\let\DAthebibliography\thebibliography
\def\thebibliography{\DeclareRobustCommand{\DA}[3]{##3}\DAthebibliography}

\usepackage[table]{xcolor}
\usepackage{graphicx}	
\usepackage{amsmath}	
\usepackage{hyperref}
\usepackage{caption}
\usepackage{subcaption}
\usepackage{comment}
\usepackage{multirow}
\usepackage{arydshln}

\newcommand{\cel}{{\sc celerite}}

\title[Fitting stochastic processes in light curves]{Modelling stochastic and quasi-periodic behaviour in stellar time-series: Gaussian process regression versus power-spectrum fitting}

\author[N. K. O'Sullivan and
S. Aigrain]{
Niamh K. O'Sullivan,$^{1}$\thanks{E-mail: niamh.osullivan@physics.ox.ac.uk}
 and Suzanne Aigrain,$^{1}$
\\
$^{1}$Denys Wilkinson Building, Department of Physics, University of Oxford, OX1 3RH, UK\\
}

\date{Accepted XXX. Received YYY; in original form ZZZ}

\pubyear{2024}

\begin{document}
\label{firstpage}
\pagerange{\pageref{firstpage}--\pageref{lastpage}}
\maketitle

\begin{abstract}
As the hunt for an Earth-like exoplanets has intensified in recent years, so has the effort to characterise and model the stellar signals that can hide or mimic small planetary signals. Stellar variability arises from a number of sources, including granulation, supergranulation, oscillations and activity, all of which result in quasi-periodic or stochastic behaviour in photometric and/or radial velocity observations. Traditionally, the characterisation of these signals has mostly been done in the frequency domain. However, the recent development of scalable Gaussian process regression methods makes direct time-domain modelling of stochastic processes a feasible and arguably preferable alternative, obviating the need to estimate the power spectral density of the data before modelling it. In this paper, we compare the two approaches using a series of experiments on simulated data. We show that frequency domain modelling can lead to inaccurate results, especially when the time sampling is irregular.  By contrast, Gaussian process regression results are often more precise, and systematically more accurate, in both the regular and irregular time sampling regimes. While this work was motivated by the analysis of radial velocity and photometry observations of main sequence stars in the context of planet searches, we note that our results may also have applications for the study of other types of astrophysical variability such as quasi-periodic oscillations in X-ray binaries and active galactic nuclei variability. 
\end{abstract}




\section{Introduction}

Stochastic processes are ubiquitous in astrophysical time-series observations. Quasi-Periodic Oscillations (QPOs) arise in accreting systems, from Active Galactic Nuclei (AGN) to compact binaries, while magnetic activity, convection and oscillations in stellar interiors all give rise to stochastic variability ranging from strongly periodic to entirely aperiodic. The resulting time-series are in general too complex to describe using parametric models, but their Power Spectral Density (PSD) can often be modelled using simple analytic functions or sums thereof. It has thus become common practice, in many astrophysical sub-fields, to model time-series observations displaying stochastic behaviour by estimating their PSDs and fitting the latter in the frequency domain. 

The functions used to model the PSD are often generalized forms of the Lorentzian or Cauchy distribution:
\begin{equation}
\label{eq:generalized_lorentzian}
    P(\nu) = \frac{\alpha}{1 + \left(\frac{\nu-\nu_0}{\gamma}\right)^\beta},
\end{equation}
where $\nu_0$ is the central frequency, $\alpha$ is the amplitude, $\beta$ is the slope and $\gamma$ is the width. (For a standard Cauchy or Lorentzian distribution normalised to unity, $\beta=2$ and $\pi \gamma = \alpha^{-1}$.) More complex models can be constructed by co-adding multiple components. 
The inferred parameters of the individual components are frequently given a physical interpretation, so it is important to understand how reliable these inferences are. Are there degeneracies between the parameters, particularly when multiple components are included, and when the number of components is not known \emph{a priori}? How robust are the parameter uncertainties derived from the fits?

While it is a very powerful and widespread approach, there are a number of drawbacks to modelling stochastic processes in the Fourier domain.
The dynamic range of PSDs is generally very large, spanning many orders of magnitude in both axes, and the number of independent estimates of the power at different frequencies is a strong function of the frequency itself, which makes fitting analytic functions to the PSD notoriously challenging. Furthermore, many real-world astrophysical time-series have irregular time-sampling, which precludes using the Fast Fourier Transform (FFT) to estimate the PSD. The Generalised Lomb-Scargle (GLS) periodogram is a popular method to approximate the PSD of irregularly sampled time-series data, but the results are highly sensitive to the window function of the observations \citet{2009A&A...496..577Z}. When the sampling is almost regular (for example for regular sampling with gaps), it is common practice to interpolate over the missing data, but that carries its own problems.

An alternative approach, which circumvents some of these issues, is to model the data directly in the time-domain, using a stochastic process model whose PSD can be written down analytically in terms of the model's parameters. Widely-used examples include (continuous or discrete) Auto-Regressive Moving Average (ARMA) models as well as certain types of Gaussian Process (GP) models (see \citet{2023ARA&A..61..329A} for a recent review of GP regression for astronomical time series). For a long time, the computational cost of this approach was prohibitive for time series consisting of $\gtrsim 10^3$ observations, but recent developments have overcome this limitation \citep[][]{celerite1, celerite2}{}{}. 

\begin{figure}
    \centering
    \includegraphics[width=\linewidth]{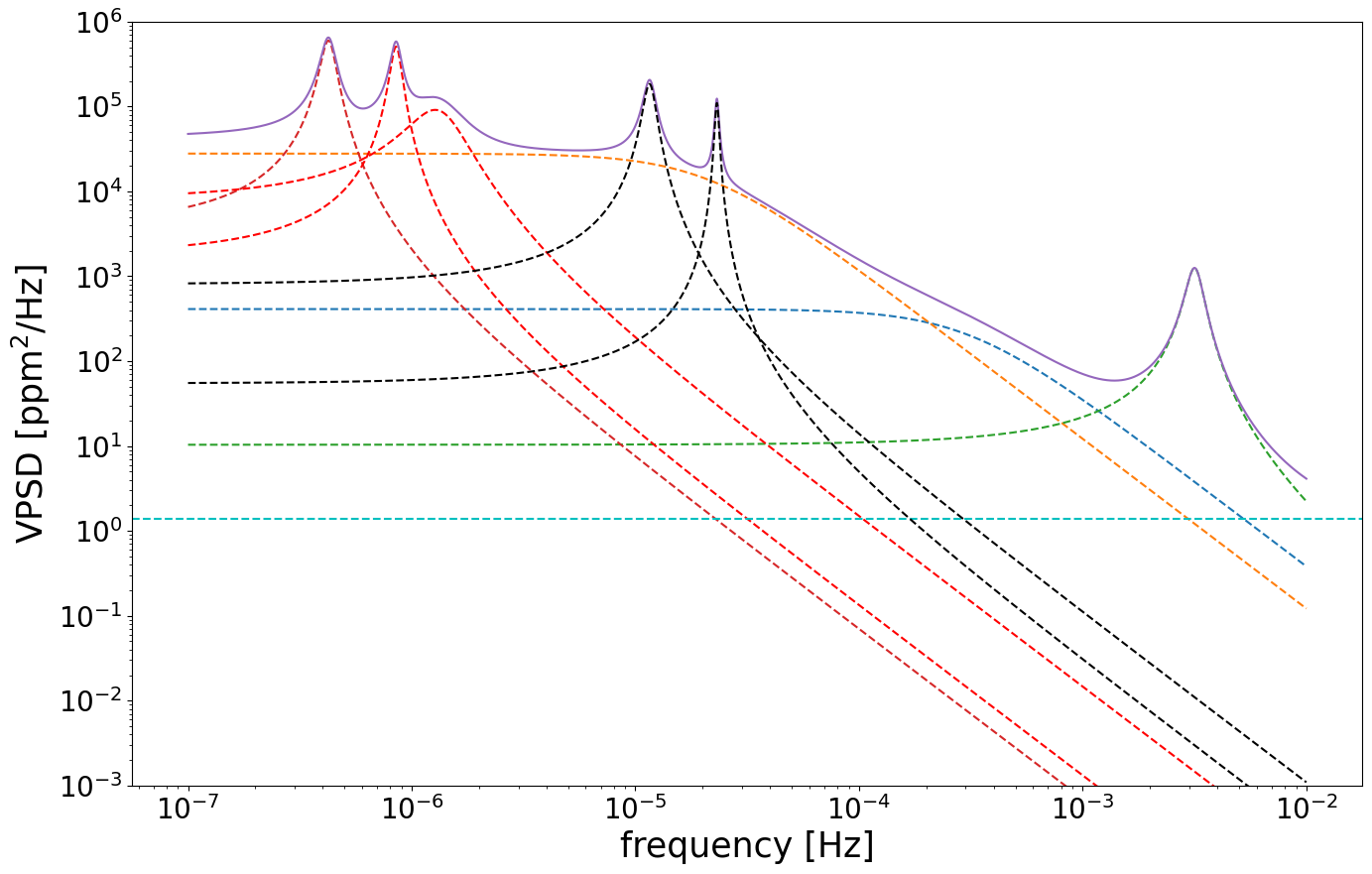}
    \caption{Composite model of the PSD of the solar background, from \citet{2023A&A...669A..39A}. The full model (solid purple line) is the sum of quasi-periodic components corresponding to the envelope of the solar oscillations (green dashed line), and to the solar rotation period and its first two harmonics  (red dashed line), aliases thereof (black dashed line), plus aperiodic components corresponding to granulation (blue dashed line) and super-granulation (orange dashed line), plus white (photon) noise (cyan dashed line).}
    \label{fig:PSD solar}
\end{figure}

In this study, we use simulations to quantify the limitations of both approaches and to ascertain when one might be preferable over the other. This work was initially motivated by recent attempts to quantify the amplitudes and timescales of different contributions to stellar `noise' in Radial Velocity (RV) planet searches, particularly the study of \citet[][hereafter A23]{2023A&A...669A..39A}, who used data from the HARPS and HARPS-North solar telescopes to reconstruct and model the PSD of solar disk-integrated Radial Velocity (RV) variability on timescales ranging from minutes to months. Their final fit to the solar PSD is reproduced in Figure~\ref{fig:PSD solar}, illustrating the complexity of the overall model, and the significant overlap between the timescales of the different components. These results are likely to have a significant impact on both the theoretical modelling of stellar (super-)granulation and the observing strategy of future Extreme Precision RV (EPRV) surveys. It is therefore important to understand how robust the fitted parameters are to the PSD-fitting methodology and to see if GP regression outperforms the `standard' approach. We also note that similar types of PSD fitting are commonplace in other fields, including asteroseismology, QPOs and AGN variability, where GP modelling is also starting to emerge as a viable alternative.

The remainder of this paper is structured as follows. We start by reviewing the common practice of fitting generalized Lorentzians to the PSD of solar variability - the so called 'solar background' in Section~\ref{sec:psd_review}. In Section~\ref{sec:gp_fitting},  we discuss GP kernels that can be used to fit the solar background in the time domain.  In Section~\ref{sec:methods}, we describe how we simulate time-series with known PSDs, and fit them them in the frequency- and in the time-domain. We then evaluate and compare the precision and accuracy of the two approaches, for different intrinsic PSDs and time-sampling properties, in Section~\ref{sec:results}. We discuss the implications of our findings and outline our future work in Section~\ref{sec:discussion}.

\section{Modelling the solar background in the frequency domain}
\label{sec:psd_review}

In this section, we review the practice of characterising stellar variability on timescales ranging from hours to months by fitting a sum of modified Lorentzian functions to the PSD of the observations. This approach was first developed by 
\citet{1985ESASP.235..199H}, 
in the context of helioseismic RV observations, and has since become widespread, being routinely applied to both solar and stellar RV and photometric observations \citep[see e.g.][]{1994ApJ...424..466G,2004A&A...414.1139A,2023A&A...669A..39A} 

\subsection{The `solar background'}
\label{sec:solar_bkg}

The term `solar background' refers to the broad spectrum of solar variability in RV or photometry, which acts as a nuisance signal in the detection and study of $p$-mode oscillations \citep[see e.g.][]{1985ESASP.235..199H}. It is modelled as the sum of up to five components, in order of decreasing frequency: the broad envelope of $p$-mode oscillations, caused by trapped sound waves in the stellar atmosphere \citep[][]{1995A&A...293...87K, 1994ApJ...424..466G}, granulation, caused by convective upflows \citep[][]{1995ApJ...444L.119R, 2008A&A...490.1143L, 1991A&A...248..245R}, super-granulation, the origin of which is still under debate (see \citealt{2010LRSP....7....2R}), the rotational modulation of active regions, caused by stellar spots rotating in and out of our field of view \citep[][]{1997ApJ...485..319S, 2010A&A...512A..39M, 2016MNRAS.457.3637H}, and long term trends caused by the star's activity cycle \citep[][]{2010A&A...511A..54S}. Beyond helio- (and astero-)seismology, this background is also an important nuisance signal that needs to be understood and mitigated for exoplanet searches, as it can hide and mimic planet signals in both Doppler and photometric stellar time-series.  

As illustrated in Figure~\ref{fig:PSD solar}, the components of the solar background have timescales ranging from minutes (in the case of $p$-modes), to hours (granulation), to days (super-granulation), to weeks or months (rotation) and in some cases decades (magnetic activity cycles). Carefully designed observing strategies can be used to minimize the impact of the higher-frequency components on RV planet searches, for example by selecting exposure times that average out the $p$-modes and spacing out consecutive exposures by more than the characteristic timescale of granulation \citep[][]{2011A&A...525A.140D}. However, the super-granulation component is particularly problematic, as it has characteristic timescales of 1--2 days \citep[][]{2015A&A...583A.118M}, which limits the frequency with which a given target can be monitored without being affected by correlated noise. \citet{2021arXiv210406072M} provides a more detailed review of the effect of solar and stellar variability on RV planet searches.

The characterisation of the components of the solar background has been traditionally done by fitting analytic functions to estimates of the PSDs of the observations. In the rest of this section, we give a brief overview of the functional forms used in the literature for modelling the different components and discuss some of the subtleties of the fitting process.

\subsection{Aperiodic Components}
\label{sec:psd_aper_functions}

\citet{1985ESASP.235..199H} introduced the notion of modelling granulation and super-granulation signals in solar RVs as a random process with exponentially decaying auto-covariance, with variance $\sigma^2$ and characteristic decay timescale $\tau$:
\begin{equation}
    \label{eq:k_exp}
    k_{\rm H}(t,t') = \sigma^2 \exp \left(-\frac{|t-t'|}{\tau}\right). 
\end{equation}
The PSD of this process is 
\begin{equation}
    \label{eq:P_H85_norm}
    P_{\rm H}(\nu) = \frac{2 \sigma^2 \tau}{1+(2 \pi\nu \tau)^2}.
\end{equation}
This is often known as a Harvey function (hence the subscript "H"). 

There is a difference of a factor 2 between the expression given by \citet{1985ESASP.235..199H} for the PSD of such a process and Equation~(\ref{eq:P_H85_norm}); we can check that the latter is correct by integrating it over the full frequency domain:
\begin{equation}
    \label{eq:V_H85}
    V_{\rm H} = \int_{-\infty}^{+\infty} P_{\rm H}(\nu) \, {\rm d} \nu = \sigma^2,
\end{equation}
satisfying Parseval's theorem. This discrepancy means that some of the literature estimates of $\sigma$ derived from PSD fits of this type may need to be adjusted by a factor $\sqrt{2}$.

In any case, later studies  \citep[e.g.,][]{1993ASPC...42..111H, 2008A&A...490.1143L, 2009A&A...495..979M, 2013ApJ...767...34K, 2014A&A...570A..41K} have shown that the power spectrum of granulation in particular is significantly steeper, and better approximated by a PSD of the form
\begin{equation}
    \label{eq:P_H93}
  P_{\rm H}(\nu) = \frac{A_{\rm H}}{1+(2 \pi \nu \tau)^4}.
\end{equation}
This is the expression used in more recent work, including A23. The difference between Equations~(\ref{eq:P_H85_norm}) and (\ref{eq:P_H93}) is illustrated on Figure~\ref{fig:PSD Forms}.
The variance of this process is
\begin{equation}
    \label{eq:V_H93}
  V_{\rm H} = \int_{-\infty}^{+\infty} P_{\rm H}(\nu) \, {\rm d} \nu = \frac{A_{\rm H}}{2 \sqrt{2} \tau},
\end{equation}
and thus depends on $\tau$ as well as $A_{\rm H}$. 

\citet{1997A&A...328..229N} provide a physical explanation for the apparently steeper power spectrum of granulation. They argue that the \citet{1985ESASP.235..199H} model corresponds to a turbulent cascade, which has power on all timescales, whereas granulation is a convection phenomenon, for which there is a minimum characteristic spatial (and thus temporal) scale, and thus a steep decay in the power spectrum.

Note that some studies, for example \citet{1993ASPC...42..111H}, let the index of the power spectrum be a free parameter, usually denoted as $c$:
\begin{equation}
    \label{eq:P_H93c}
    P_{\rm H}(\nu) = \frac{A_{\rm H}}{1+(2 \pi \nu \tau)^c},
\end{equation}
but this tends to result in fitted values in the range $3.5 \le c \le 5.5$, consistent with $c=4$ within the uncertainties. 

For lower-frequency components such as super-granulation (in RV) or faculae (in photometry, \citealt{2013ApJ...767...34K}), even steeper PSD indices ($c \sim 6$) are sometimes reported, but it is not clear how robust those estimates are, as the high-frequency tails of these components overlap strongly with the granulation component.

\subsection{Periodic Components}
\label{sec:psd_per_functions}

To model (quasi-)periodic components, for example the envelope of the $p$-modes, \citet{1993ASPC...42..111H}  and later \citet{2008A&A...490.1143L} used a modified Lorentzian function of the form:
\begin{equation}
    \label{eq:P_LL1}
    P_{\mathrm{L}}(\nu)= A_{\mathrm{L}} \left( \frac{\nu}{
  \nu_0}\right)^b \left[ \frac{\Gamma^2}{(\nu-\nu_0)^2 + \Gamma^2} \right]^c,
\end{equation}
where $\nu_0$ is the central frequency and $A_{\mathrm{L}}$ is the power at $\nu=\nu_0$. To reduce the number of free parameters, many recent studies that use this type of periodic model, including A23, fix $b=0$ and $c=1$, so that 
\begin{equation}
    \label{eq:P_LL2}
    P_{\mathrm{L}}(\nu)= A_{\mathrm{L}} \frac{\Gamma^2}{(\nu-\nu_0)^2 + \Gamma^2}, 
\end{equation}
in which case $\Gamma$ is the half-width at half maximum (HWHM). 
The variance of this process is:
\begin{equation}
    \label{eq:V_LL2}
    V_{\mathrm{L}}(\nu)= \int_{-\infty}^{\infty} P_{\mathrm L}(\nu) \, {\rm d}\nu = \pi \, A_{\rm L} \, \Gamma.
\end{equation}
Once again, it depends not only on the amplitude but also on the width (or damping time-scale).

Some other recent studies, for example \citet{2014A&A...570A..41K}, use a Gaussian instead of a Lorentzian function to model the envelope of the $p$-modes:
\begin{equation}
    \label{eq:P_LG}
   P_{\mathrm{L}}(\nu)= A_{\mathrm{L}} \exp \left[ - \left(\frac{\nu-\nu_0}{\sigma} \right)^2 \right], 
\end{equation}
where in this case $\sigma$ is the standard deviation of the Gaussian (not to be confused with the standard deviation of the process). 
This is more in line with the standard asteroseismology practice for measuring $\nu_{\rm max}$, the peak frequency of the stellar $p$-modes, which is then used in scaling relations to determine the stellar mean density \citep[see e.g.][]{2013ARA&A..51..353C}. 

Neither of the above formulations have a particularly clear physical motivation, however; their use is justified empirically rather than theoretically.

\subsection{PSD estimation and fitting}
\label{sec:psd_how}

For regularly sampled observations, PSD estimation is straightforward using the Fast Fourier Transform (FFT). For irregularly sampled time-series, such as ground-based RV observations, the Lomb-Scargle (LS) periodogram \citep{1976Ap&SS..39..447L,1982ApJ...263..835S} and later derivatives thereof are the PSD estimation method of choice in the astronomical community. It is worth noting that care is needed when selecting the normalisation of the LS periodogram so that it gives results that are directly comparable to the FFT-estimated PSDs. See \citet{2018ApJS..236...16V} for a detailed discussion of the LS periodogram and its interpretation.

The fitting process itself is often challenging, owing to the wide dynamic range covered by the PSD in both frequency and power density, the difficulty in assigning relative weights to the individual PSD samples, and the overlapping nature of the background components, which leads to significant degeneracies between their parameters. Here we describe the procedure used by A23, which is fairly representative of previous studies. 
They estimate the PSD using the Generalised Lomb-Scargle (GLS) periodogram \citep{2009A&A...496..577Z}, which is then binned to a regular grid in log frequency space, to give the low-frequency components adequate weight in the fit. As previously mentioned, the slope of the individual background components is fixed to improve convergence and reduce degeneracies between the components. Any white noise in the original time-series translates to a constant lower envelope in the PSD, which is incorporated as a constant term in the fit. The fitting itself is done by minimizing the sum of squared residuals using the Levenberg-Marquart algorithm, which also provides estimates of uncertainties on the fitted parameters using the curvature of the metric around the optimum. 

The model of A23 (illustrated in Figure~\ref{fig:PSD solar}), includes aperiodic terms (Equation~\ref{eq:P_H93}) to represent granulation and super-granulation, and Lorentzian periodic terms (Equation~\ref{eq:P_LL2}) to represent the envelope of the $p$-modes and signals associated with the rotational modulation of active regions. For the latter, they include terms at the fundamental and first two harmonics of the solar rotation period, as well as 1-day aliases of the fundamental and first harmonics. The need to explicitly include terms to model the aliases is another drawback of modelling the data in the frequency domain.

\begin{figure}
    \centering
    \includegraphics[width=0.48\textwidth]{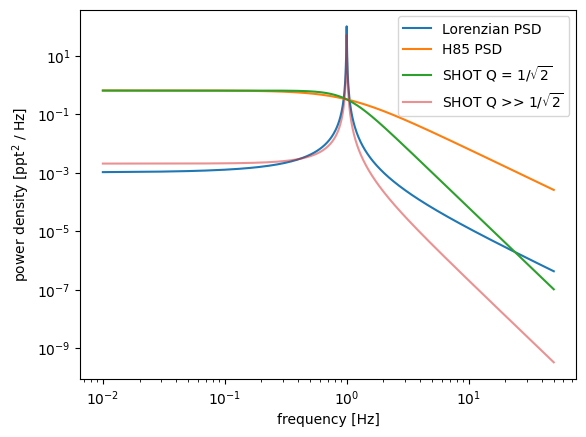}
    \caption{Comparison between the different types of PSD functions used in this paper. The orange and green curves respectively show the aperiodic Lorentzian PSD of Equation~(\ref{eq:P_H85_norm}) and the steeper-sloped version of Equation~(\ref{eq:P_H93}), while the blue and pink curves respecitvely show the periodic Lorentzian PSD of Equation~(\ref{eq:P_LL2}), and the high-$Q$ SHO term PSD of Equation~(\ref{eq:PSD_SHO}).}
    \label{fig:PSD Forms}
\end{figure}

\section{Modeling the solar background in the time domain using GPs}
\label{sec:gp_fitting}

GP regression is a natural alternative to PSD fitting when modelling stochastic or quasi-periodic processes in time-series data. Indeed, stochastic process models provided the original motivation for the choice of function used to fit the solar background PSD as described in Sections~\ref{sec:psd_aper_functions} and \ref{sec:psd_per_functions}. It is thus natural to ask whether GP regression could be used to model the same signals directly in the time domain, side-stepping some of the implementation challenges alluded to in Section~\ref{sec:psd_how}. In particular, the \cel\ package (\citealt{celerite1}, hereafter F17; \citealt{celerite2}) provides an efficient implementation of GP regression for certain classes of covariance functions that are particularly appropriate to model stellar signals, and scale well to large datasets. 

\subsection{The \cel\ SHO term}

A generic \cel\ model is constructed by adding or multiplying together basic building blocks (known as `terms') with covariance 
\begin{equation}
    \label{eq:cel_psd}
    k_{\rm cel}(t,t') = a \exp (-c |t-t'|),
\end{equation}
where $a$ and $c$ are complex numbers. These individual terms can then be combined additively or multiplicatively to construct more complex models. 

In particular, the \cel\ package provides a family of kernels that approximate the behaviour of a harmonic oscillator and are particularly appropriate to model a wide range of stellar signals, from aperiodic to strongly periodic. This is implemented in the form of the `Simple Harmonic Oscillator` (SHO) term, controlled by 3 parameters: an amplitude $S_0$, a characteristic (undamped) angular frequency $\omega_0$, and a quality factor $Q$. The PSD of the SHO term\footnote{Note that Equation~(\ref{eq:PSD_SHO_Qs2}) was obtained by substituting for $\omega= 2\pi \nu$ in Equation~(20) of F17, then mulitplying it by a factor $\sqrt{2\pi}$. This is to account for the fact that F17 defined the Fourier transform to be unitary in angular frequency units rather than in natural frequency units, resulting in a different normalisation constant for the forward transform.} is:
\begin{equation}
    \label{eq:PSD_SHO}
    P_{\rm SHO}(\nu) = \frac{2 S_0 \, \nu_0^4}{(\nu^2-\nu_0^2)^2+ \nu^2 \nu_0^2 / Q^2}.
\end{equation}
Around $\nu \approx \nu_0$, this PSD approximates a standard Lorentzian or Cauchy distribution, but it falls more steeply (with index 4) at high frequencies. 

As of \href{https://celerite2.readthedocs.io/}{version 2} of the \cel\ package, an alternative parametrisation is also available, where the user can specify the period $\rho$, damping timescale $\tau$ and standard deviation $\sigma$ of the process instead of the angular frequency, amplitude and quality factor:
\begin{equation}
    \label{eq:SHO_altpar}
    \rho \equiv 2 \pi / \omega_0, ~~~ \tau \equiv 2 Q / \omega_0 \text{~~~and~~~} \sigma \equiv \sqrt{S_0 \omega_0 Q}.
\end{equation}
This alternative parametrisation matches the variable names in Sections~\ref{sec:psd_aper_functions} and \ref{sec:psd_per_functions}, making the relationship between the two classes of models more readily apparent.

\subsection{Aperiodic components}

Clearly, it is straight-forward to reproduce the PSD of Equation~\ref{eq:P_H85_norm} using a \cel\ term with real-valued $a$ and $b$. This is also the limiting behaviour of the SHO term when $Q=1/2$ and $\omega_0 \to 0$.

To reproduce the steeper slope at high frequencies used by more recent studies, however, 
a natural choice is the SHO term with $Q = 1 / \sqrt{2}$. In this case, the PSD of the SHO term becomes:
\begin{equation}
    \label{eq:PSD_SHO_Qs2}
    P_{1/\sqrt{2}}(\nu) = \frac{2 \, S_0}{1+(\nu/\nu_0)^4}.
\end{equation}
This is exactly equivalent to Equation~\ref{eq:P_H93} if $\nu_0 \equiv 1/2 \pi \tau$ and $2 \, S_0 \equiv A_{\rm H}$. 

\subsection{Periodic terms}

The PSD of a generic \cel\ term with real-valued $a$ but complex-valued $c$ is (adapting Equation 11 of F17):
\begin{equation}
  \label{eq:P_C}
  P_{\mathrm{C}}(\nu) 
  = \frac{a}{c} \, \left[ 
  \frac{1}{1+\left( \frac{ 2 \pi \nu -d}{c} \right)^2} + 
  \frac{1}{1+\left( \frac{ 2 \pi \nu +d}{c} \right)^2}
  \right].  
\end{equation}
If we set $a \equiv 2 \pi A_{\mathrm{L}} \Gamma$, $c \equiv 2 \pi \Gamma$ and $d \equiv 2 \pi \nu_0$, we obtain
\begin{equation}
  P_{\mathrm{C}}(\nu) = A_{\rm L} \, \left[ 
  \frac{\Gamma^2}{\left( \nu - \nu_0 \right)^2 + \Gamma^2} + 
  \frac{\Gamma^2}{\left( \nu + \nu_0 \right)^2 + \Gamma^2}
  \right],
\end{equation}
which is the sum of two of the Lorentzian functions in Equation~(\ref{eq:P_LL2}), with characteristic frequencies $\pm \nu_0$.  This is a fairly close match to the periodic components used by A23, but not an exact one, because of the negative frequency term.

On the other hand, we know that $p$-modes are stochastically driven oscillations. Therefore, it makes sense to model them as such, using an SHO term with large $Q$, with PSD given by Equation~(\ref{eq:PSD_SHO}). If we set $S_0 = A_{\rm L} \, \Gamma^2 / \nu_0^2$ and $Q = \nu_0 / 2 \Gamma$, the resulting process will have the same variance and coherence time as the one defined by the Lorentzian PSD of Equation~(\ref{eq:P_LL2}). As Figure~\ref{fig:PSD Forms} shows, the PSDs of these two processes look very similar in the vicinity of the peak at the characteristic frequency, but diverge away from it, particularly at high frequencies, where the SHO term gives rise to a steeper power-law slope, with index 4 rather than 2.

\subsection{The \cel\ rotation term}
\label{sec:rotation_term}

The PSD of solar and stellar variability often displays peaks at the first harmonic of the rotation period as well as the fundamental, which need to be modelled explicitly to obtain a good fit. For example, A23 included periodic components at the first two harmonics of the solar rotation period (see Section~\ref{sec:psd_how}) in their final fit. \citet{Foreman-Mackey2021} recommends modelling stellar rotation signals using the sum of two under-damped SHO terms: one at the fundamental period, and one at its first harmonic. This is implemented in \href{https://celerite2.readthedocs.io/}{version 2} of the \cel\ package as the `rotation term`.

The rotation term is controlled by 5 parameters: the standard deviation $\sigma$ of the process, the period $P$, two parameters controlling the quality factors of the fundamental and first harmonics (these are specified in such a way that the fundamental has quality factor $Q_1>1/2$ and the harmonic has $Q_2>Q_1$), and one parameter corresponding to the amplitude of the first harmonic term relative to the fundamental.

\section{Test methodology}
\label{sec:methods}

In this section, we describe how we simulate time-series datasets with known PSDs including either periodic terms, aperiodic terms or a mixture of the two, and model them both in the frequency domain using analytic PSD functions, and in the time-domain using a GP, to compare the two approaches directly. We restrict ourselves to relatively simple processes with one or two components but also test the impact of irregular time-sampling.

\subsection{Simulating the time-series}

\begin{figure*}
    \centering
    \includegraphics[width=0.98\textwidth]{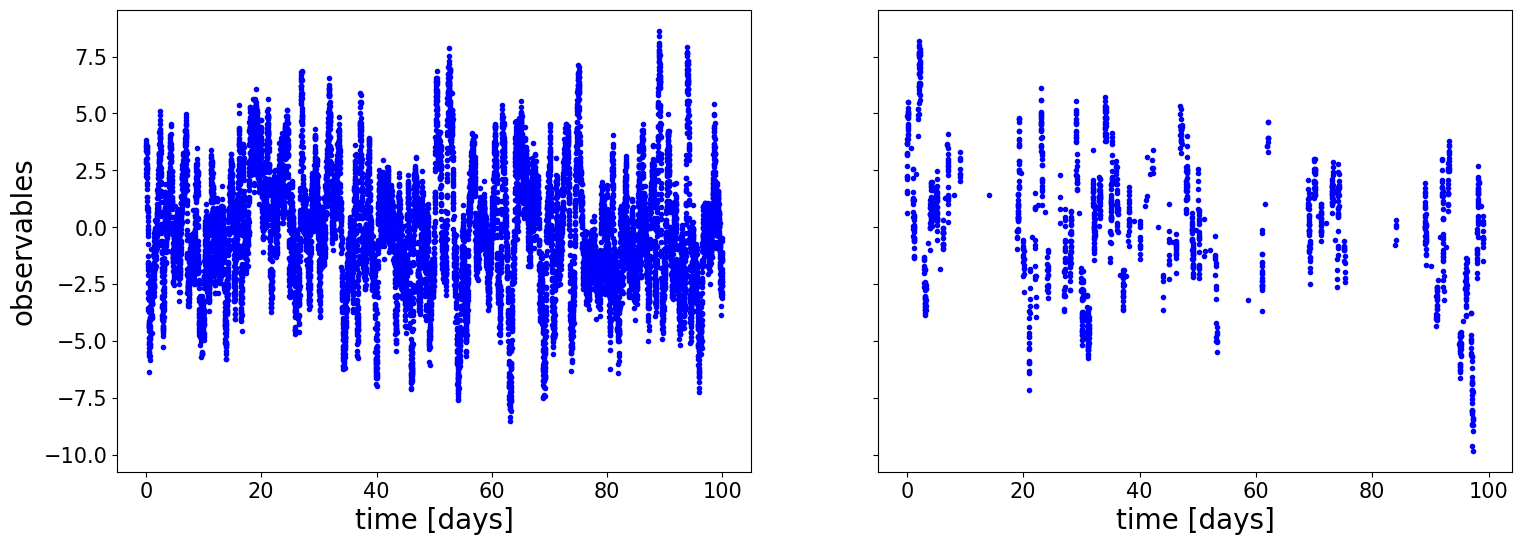}
    \caption{Left: Example of a simulated, noise-free, regularly sampled super-granulation time series created using a \cel\ SHO term GP with $Q=1/\sqrt{2}$, $S_0 = 2$ and $\nu_0 = \frac{5}{2\pi}$\,cycles/day. Right: noisy, irregularly sampled time-series generated for the same process with time-stamps from the HARPS-N solar telescope.}
    \label{fig:aperiodic_TS}
\end{figure*}

We first simulated a set of regularly spaced time stamps lasting 288 days with 100 points per day, corresponding to a 5 minute cadence. This corresponds to a Nyquest frequency of $452.38\,{\rm day}^{-1}$. We then simulated observations with a known PSD by drawing samples from a \cel\ GP. For aperiodic components, we use an SHO term with $Q=1/\sqrt{2}$, the PSD of which is given by Equation~\ref{eq:PSD_SHO_Qs2}. For periodic terms, we use an SHO term with large $Q$, the PSD of which is given by Equation~(\ref{eq:PSD_SHO}). An example of such a simulated time-series for an aperiodic component with $Q=1/\sqrt{2}$, $S_0=2$ (in arbitrary units) and $\omega_0=5$\,radians/day (such as might be expected for a super-granulation signal) is shown in the left-hand panel of Figure~\ref{fig:aperiodic_TS}.

Finally, white noise with a standard deviation of $\sigma_{\rm w}=0.5$ was added to the simulated time-series to represent measurement uncertainties.

We later investigate the impact of changing the integration time by binning the simulated time-series, and of the baseline by truncating them.

To produce time series with irregular time-sampling, such as might be expected for ground-based RV observations, we used a representative set of time stamps from the publicly available dataset obtained by the HARPS-N solar telescope between 2015 and 2018 \citep[][]{2021A&A...648A.103D}. The HARPS-N solar telescope observes the Sun at 5 minute cadence for several hours every day, with occasional interruptions caused by weather and technical issues.We take a 100 day subset of the HARPS-N time stamps and then simulated observations in the same way as described above for the regularly sampled case. As the HARPS-N data has a roughly 5 minute cadence, we expect the Nyquist frequency to be similar to the regular time sampling case. An example of the resulting irregularly sampled time-series is shown in the right-hand panel of Figure~\ref{fig:aperiodic_TS}.

Note that all the variances, standard deviations and amplitudes discussed in this section are in arbitrary units, but could represent normalised flux units (such as parts-per-thousand or parts-per-million) or relative RV units (such as m/s or cm/s).

\subsection{Estimating PSDs}

\begin{figure*}
    \centering
    \includegraphics[width=0.98\textwidth]{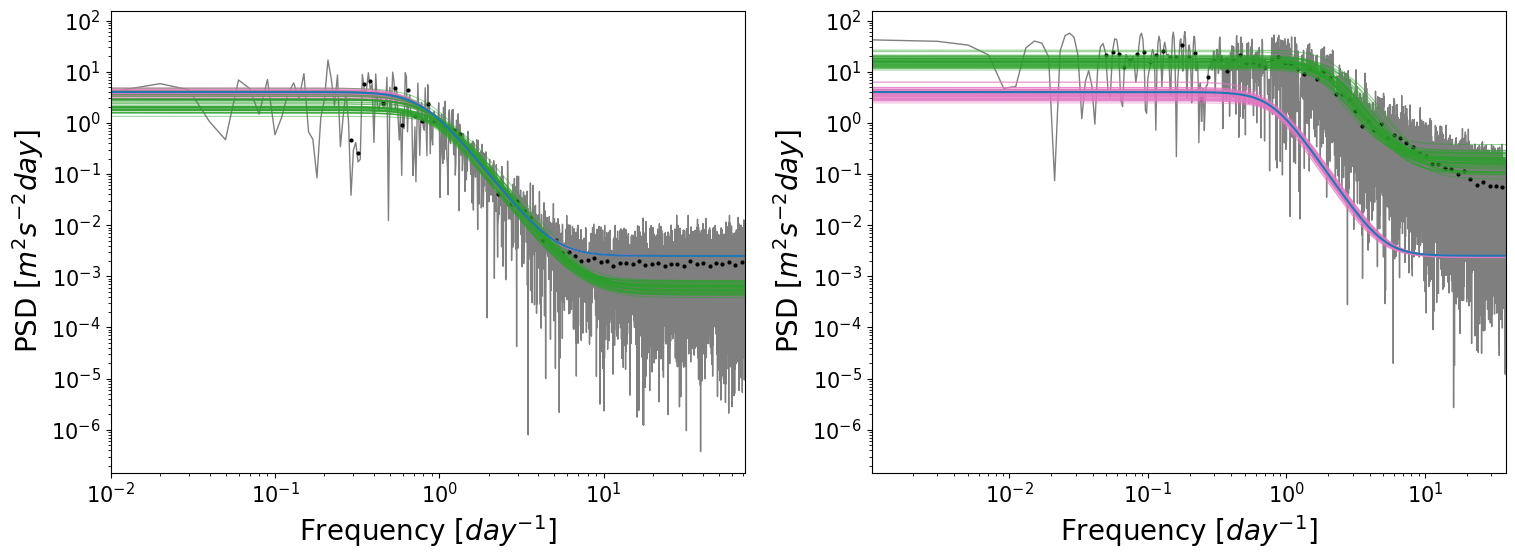}
    \caption{PSD estimates and fits for a single, aperiodic component with $S_0 = 2$ and $\nu_0 = 5/2\pi$, with regular sampling (left) and irregular sampling (right). The blue line shows the true PSD of the process used to simulate the data, which is the same in both panels. The grey line shows the PSD estimated using the FFT (left) and GLS (right), respectively, while the black dots show the binned version used in the PSD fit. The green lines show posterior samples from the PSD-fitting MCMC chain, and the pink lines show the same but from the GP fit.  Fitted values can be found in Tables \ref{tab:regular all cases} and \ref{tab:irrregular all cases}. }
    \label{fig:aperiodic_PSD}
\end{figure*}

For regularly sampled time-series, we used the {\sc numpy} implementation of the Fast Fourier Transform (FFT) to estimate the Discrete Fourier Transform of the observations. The PSD is then given by the squared modulus of the DFT, divided by the frequency step $1/T$ where $T$ is the duration of the simulated time array. The left panel of Figure~\ref{fig:aperiodic_PSD} shows the FFT-estimated PSD for the simulated time-series shown in the left hand panel of Figure~\ref{fig:aperiodic_TS}.

For irregularly sampled time-series, we used the {\sc astropy} implementation of the Generalised Lomb Scargle GLS) periodogram \citep{2009A&A...496..577Z} with the normalisation parameter set to `psd'  to estimate the PSD of the observations on a regular grid of frequencies between $\nu=0$ and $\nu=0.1$\,cycles/min (corresponding to the approximate effective Nyquist frequency of the HARPS-N solar observations), with frequency step $1/T$ where $T$ was the duration of the observations. We found it was necessary to divide the output of the periodogram code by the frequency step and by the number of data points to obtain a properly normalised PSD. As the right-hand panel of Figure~\ref{fig:aperiodic_PSD} shows, the PSDs obtained in this manner for irregularly sampled time-series can be strikingly different to the true model used to generate the observations, especially at high frequencies. Naturally, these differences will translate to differences between the true and fitted parameters, highlighting the limitations of PSD fitting for irregularly sampled observations.

\begin{figure*}
    \centering
    \includegraphics[width=0.98\textwidth]{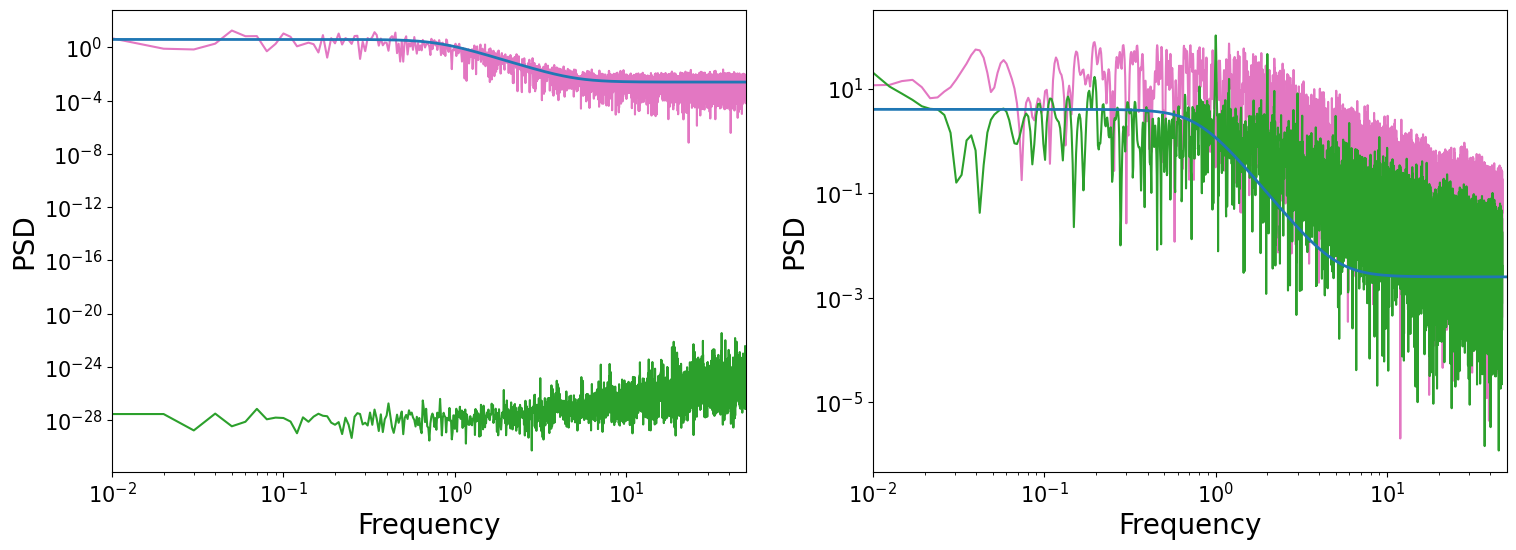}
    \caption{Impact of the window function on the PSD of a for a single, aperiodic component with $S_0 = 2$ and $\nu_0 = 5/2\pi$, with regular sampling (left) and irregular sampling (right). The PSD estimate of the window function is shown in green, the estimated PSD of the simulated data is shown in pink, while the true PSD is shown in blue}
    \label{fig:Window Function}
\end{figure*}

Although the Lomb Scargle or GLS periodograms are widely used to evaluate the PSD of irregularly sampled time-series, they do not explicitly account for the window function of the data, which can significantly alter the PDS estimate. A set of finite-duration, irregularly sampled observations of a continuous function $f(t)$ can be represented as the convolution of $f(t)$ and a window function $w(t)$, which is equal to one during each observation and zero elsewhere. If the duration of the observations is short compared to the time-scale of variations in $f(t)$, the window function can be approximated by a sum of delta functions located at the times of the observations, $\{t_n\}$. The Fourier transform of the observations is then the convolution of the Fourier transform of $f(t)$, $F(\nu)$, and of the Fourier transform of the window function, $W(\nu)$. We can gain some insight into the impact of the window function on the estimated PSD by computing the PSD of the window function, which (for instantaneous observations) is given by \citep[see e.g.][]{2018ApJS..236...16V}:
\begin{equation}
    \label{eq:WF_PSD}
     P_W(\nu; \{t_n\}) = \left | \sum_{n=1}^{N} i\sin{2 \pi \nu t_n} + \cos{2 \pi \nu t_n} \right | ^2. 
\end{equation}
To illustrate this impact, we show in Figure~\ref{fig:Window Function} the PSD of a single aperiodic term estimated from data with both regular and irregular time-sampling, together with the PSD of the window function for both cases. This highlights how the window function dominates the irregularly sampled PSD, which differs substantially from that of the true PSD. Note that the peaks in the window function PSD (corresponding to periodicities in the observation times, e.g. 1 day) do not appear clearly in the PSD of the observations, because the PSD of the true signal, which it is convolved with, has a broad spectrum (if the true signal was strongly harmonic, this would not be the case). Ideally one would seek to deconvolve the PSDs of the signal and the window function, but this is not possible in practice because the window function is zero for most values of $t$ \citep{2018ApJS..236...16V}. As a result, most authors that fit PSD models for irregularly sampled data usually include a constant term to account for the extra white noise induced by the window function but do not explicitly account for its structure beyond that. 

This is one reason why we might expect that modelling stochastic processes in the time domain should be particularly advantageous over doing so in the frequency domain for irregularly sampled observations. 


\subsection{Fitting PSDs}
\label{sec:fit_psd}

\begin{figure*}
     \centering
     \begin{subfigure}[b]{0.48\textwidth}
         \centering
         \includegraphics[width=\textwidth]{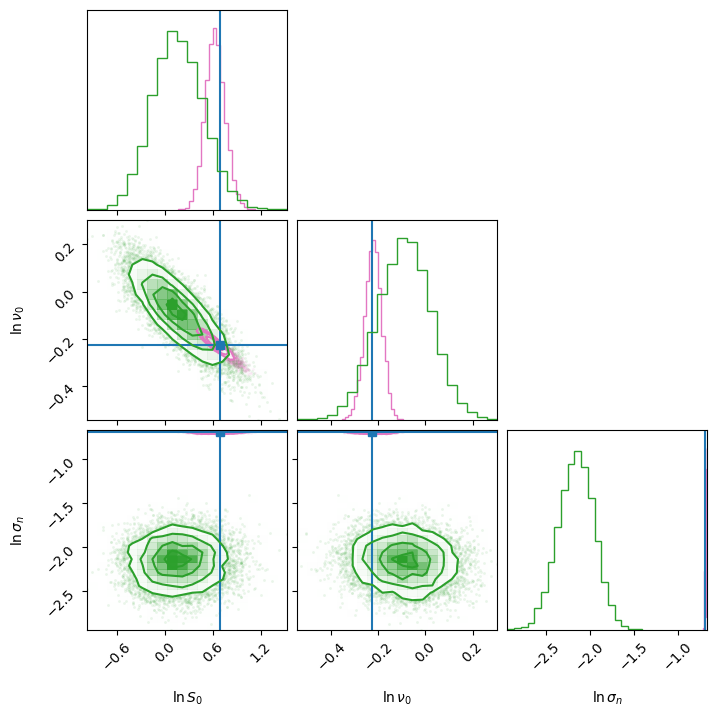}
     \end{subfigure}
     \hfill
     \begin{subfigure}[b]{0.48\textwidth}
         \centering
         \includegraphics[width=\textwidth]{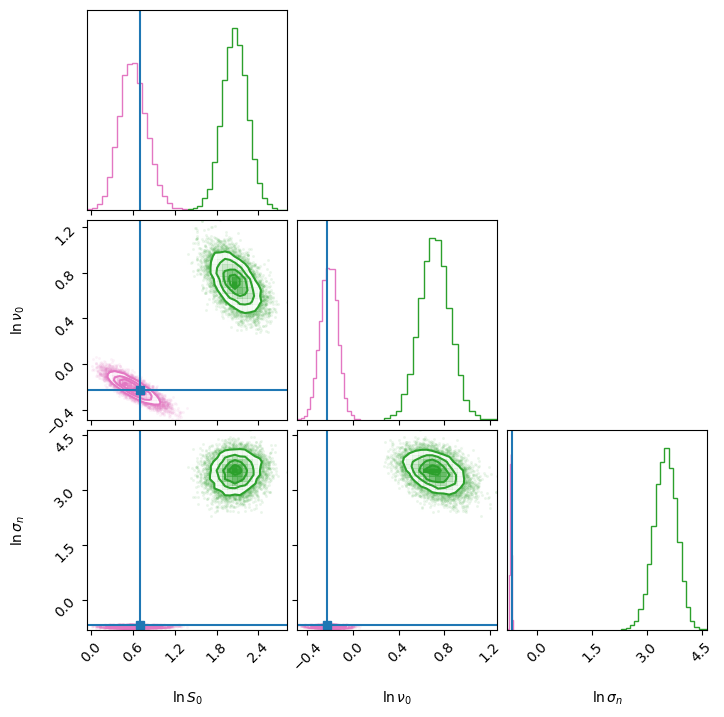}
     \end{subfigure}
     \caption{MCMC posterior distribution plots for the case of a single, aperiodic component with $S_0 = 2$ and $\nu_0 = 5/2\pi$, with regular sampling (left) and irregular sampling (right). These correspond to the time-series shown in Figure~\ref{fig:aperiodic_TS} and to the PSDs shown in Figure~\ref{fig:aperiodic_PSD}. The 1-D posterior distributions for each parameter, marginalised over all the other parameters, are shown by the histograms in the diagonal panels, in green for the PSD fits and in pink for the GP fits, with the true parameter values indicated by the blue lines. The 2-D posteriors are shown in the off-diagonal panels using the same colour-coding, with 1-, 2- and 3-$\sigma$ contours.   Fitted values can be found in Tables \ref{tab:regular all cases} and \ref{tab:irrregular all cases}.}
     \label{fig:aperiodic_corner}
\end{figure*}

Once the PSD corresponding to a given observation has been estimated, we proceed to fit it, using a PSD model function corresponding to the true PSD of the GP used to generate the data in the first place (namely Equation~\ref{eq:PSD_SHO_Qs2} for aperiodic components, and Equation~\ref{eq:PSD_SHO} for periodic components). The parameters of the model for each component are $\ln S_0$ and $\ln \omega_0$, plus $\ln Q$ for the periodic components only. Fitting for the logarithm of the parameters improves convergence and ensures that they always remain positive. Additionally, a constant term, $\sigma_{\rm w}$, (also fit in log space) is included in the fit to account for white noise and any under-sampled signals. The constant added to the PSD model is set to $\sigma_{\rm w} / 2 T$, where $T$ is the total duration of the time-series.

Following A23, we bin the PSD into 100 equally spaced bins in log frequency space before fitting, except when  working with periodic terms, where we found it necessary to increase the number of bins to 200 in order to properly resolve the PSD peak. We use only bins containing a minimum of 3 samples (corresponding to frequencies above 0.28 cycles/day), and discard the PSD estimates at lower frequencies. We note that binning PSDs before fitting is normally done in part to ensure that the PSD samples used in the fit are approximately uncorrelated. This would require at least 10, rather than 3, samples per bin. However, this would translate to a minimum frequency of $geq 1$ cycle/day, which is very close to the typical timescale for super-granulation in Sun-like stars, and would therefore preclude the extraction of any information about super-granulation timescales from our simulated time-series. Our decision to use bins with at least $N_{\rm min}=3$ samples represents a compromise between maximising the frequency range and minimizing correlation between bins. All bins we assigned equal weight in the fits. Again, this follows the procedure described in A23, where no mention is made of assigning weights or uncertainties to the binned PSD samples.

We perform the fit in log-space, i.e. we minimize the sum of the squared differences between the logarithm of the binned PSD and that of the model function, and assign the same weight to all the binned samples. We also carried out some tests where the fits were performed in linear space and the samples were weighted according to the standard deviation in each bin. However, we found that standard deviation of the PSD samples in each bin systematically underestimates the uncertainty on the binned PSD samples, resulting in poor fits and unrealistically small uncertainties on the fit parameters.

We use Markov Chain Monte Carlo (MCMC) to sample the joint posterior probability distribution over the parameters of the fit, using version 3 of the {\sc emcee} package \citep{2013PASP..125..306F,2019JOSS....4.1864F}. The walkers are initialised in a tight Gaussian ball (with standard deviation 0.01\,dex) around a local optimum found using the {\sc minimize} function in {\sc scipy}'s {\sc optimize} module. Log-uniform priors are used for all the parameters, within the interval $[-11;11]$ in all cases except for the white noise standard deviation $\log \sigma_{\rm w}$, which we restrict to the interval $[-2;2]$. 
The number of walkers is set to 4 times the number of parameters for each fit, and the MCMC chains are run for 10\,000 steps. 

We use {\sc emcee}'s built-in functionality to estimate the auto-correlation length of the chains, in order to assess convergence and select an appropriate burn-in and thinning factor. Provided that the longest auto-correlation length estimate across all parameters is $\tau_{\rm max} \le 200$ steps, i.e.\ less than a $50^{\rm th}$ of the chain, we consider the auto-correlation time estimates to be reliable and the chains well-converged. We then discard the first $3 \tau_{\rm max}$ samples as part of the burn-in phase, and thin the remainder of the chains by a factor $\tau_{\rm max}/4$. In the few cases where the chains do not converge in 10\,000 steps (predominantly when using the PSD-fitting method for multi-component models), we arbitrarily set the burn-in time to 1000 steps and the thinning factor to 30, but note that the results should be treated with caution. 

For illustrative purposes, we evaluated the PSD model for 50 random samples from the truncated, thinned MCMC chains, which are shown in green on the PSD plots in this paper (such as Figure~\ref{fig:aperiodic_PSD}). We also use the {\sc corner.py} package \citep{corner} to display 1- and 2-dimensional posteriors for each fit and compute the median estimates of each parameter and the associated 1-$\sigma$ uncertainties. Figure~\ref{fig:aperiodic_corner} shows the corner plots obtained in this manner for the time-series shown in Figure~\ref{fig:aperiodic_TS}, for which the PSDs are shown in Figure~\ref{fig:aperiodic_PSD}.

\subsection{GP-modelling}

We also modelled the simulated time-series directly in the time domain, using a GP model of the same form as the one used to generate them. The fit parameters, priors and details of the optimization and posterior sampling procedures were identical to those described in Section~\ref{sec:fit_psd}, the only difference being that the figure of merit of the fit was the GP likelihood, rather than the $\chi^2$ of the PSD fit. 

The results of the GP fits are shown in pink on the PSD and corner plots (such as Figures~\ref{fig:aperiodic_PSD} and \ref{fig:aperiodic_corner}), while the results of the PSD fits are shown in green on the same figures, and the true model in blue.

\section{Results}
\label{sec:results}

We are now ready to compare the results obtained for these simulated time-series with both the PSD-fitting and the GP-modelling approach, varying the type of model used and the time-sampling.

\subsection{Single-component case}

We start with the simplest case, namely the single, aperiodic, (super-)granulation-like component illustrated in Figures~\ref{fig:aperiodic_TS}, \ref{fig:aperiodic_PSD} and \ref{fig:aperiodic_corner}. The fitted parameters for all the tests described in this section are reported in Tabels \ref{tab:regular all cases} and \ref{tab:irrregular all cases}.

Even in the regularly-sampled case, the parameter estimates derived via PSD fitting already deviate from the true values at the 2--3$\sigma$ level. This is not unexpected for the white noise component: the finite duration of each observation and of the total time-series result in additional white noise in the PSD coming from the window function. However, it is concerning to see a significant deviation for the other parameters, even in this idealised case. The discrepancy between injected and recovered parameters is even starker in the irregular case, as expected due to the more complex window function. The PSD fits are also quite sensitive to somewhat arbitrary modelling choices such as the number of PSD bins used.

\begin{figure*}
    \centering
    \includegraphics[width=0.98\textwidth]{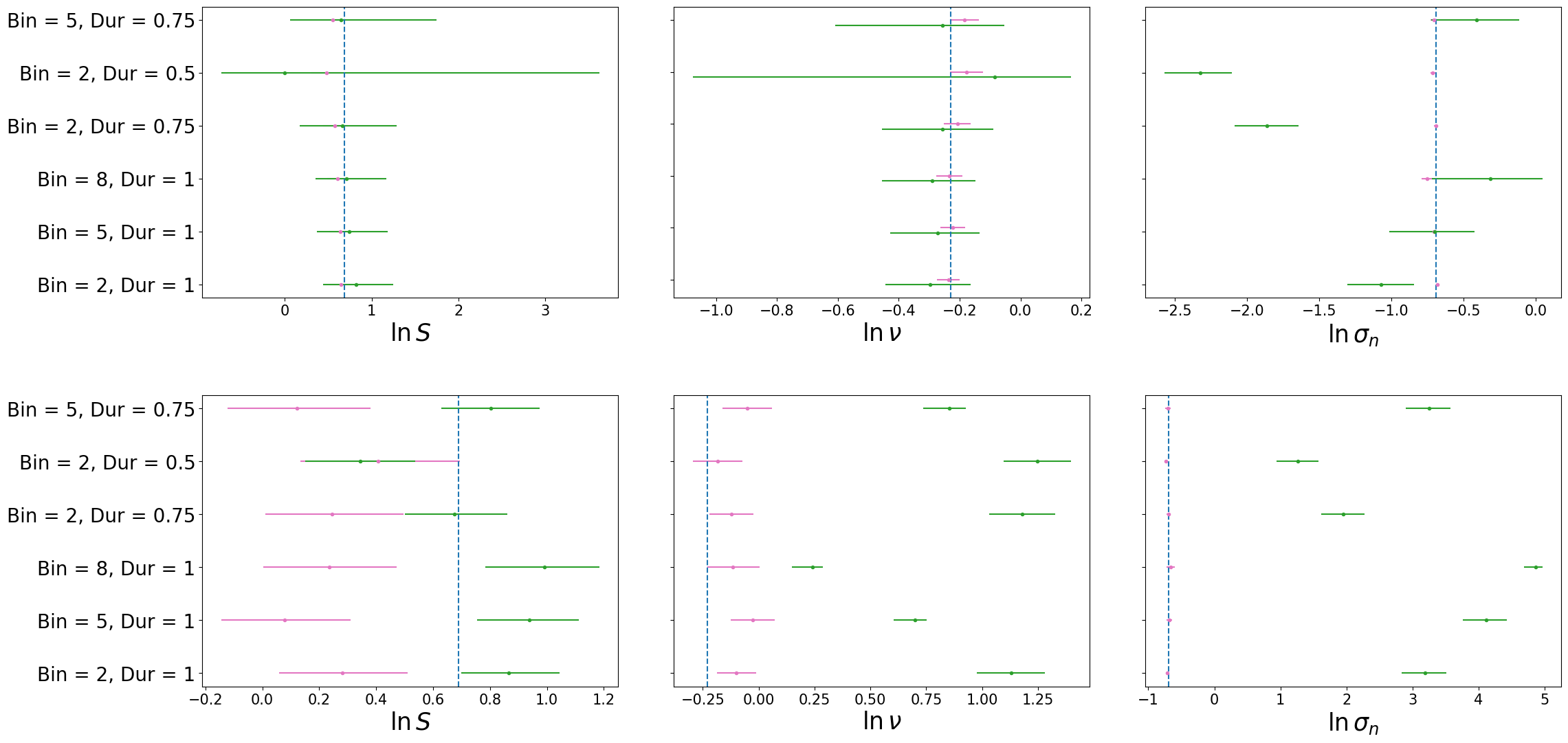}
    \caption{Recovered versus true parameters for a single aperiodic component with regular time sampling (top row) and irregular sampling (bottom row), as the binning is increased and the duration of the time-series is truncated. Green symbols show the results of the PSD fits, while pink symbols show those of the GP fits. The true parameter values are shown by the vertical blue lines. }
    \label{fig:downgrading binning compariosn}
\end{figure*}

We then explored the effect of both increasing the exposure times (i.e. binning the time-series) and reducing the duration (truncating the time-series) to test the sensitivity of the results to these factor. As shown in Figure \ref{fig:downgrading binning compariosn}, this  degraded the accuracy of the results further, especially in the irregular case, where all parameters were overestimated, even when the `knee' of the PSD appears well sampled by eye. In the regular case, the PSD fits, while generally consistent with the the true value, have larger uncertainties associated with them.

By contrast, the GP-modelling results are both more precise and consistent with the truth in all these cases, though the precision worsens (as expected) when the time-series is binned, truncated or irregularly sampled. We repeated this test multiple times with a range of $\omega_0$ and $S_0$ values and found that the results were substantially the same. 

\begin{figure*}
     \centering
        \includegraphics[width=\textwidth]{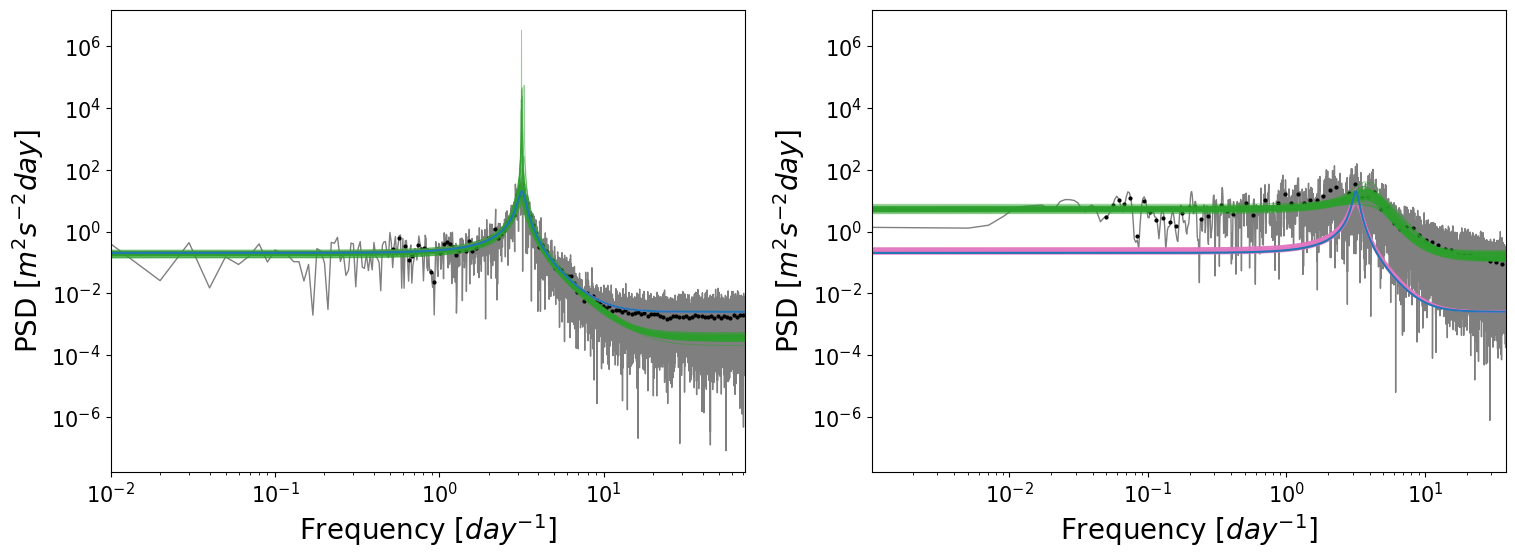}
     \caption{PSD estimates and fits for a single, periodic component with $Q=10$,  $S_0 = 0.1$ and $\nu_0 = 20/2\pi$, with regular sampling (left) and irregular sampling (right). The colour coding is the same as Figure \ref{fig:aperiodic_PSD}. Fitted values can be found in Tables \ref{tab:regular all cases} and \ref{tab:irrregular all cases}.
     }
     \label{fig:periodic_PSD}
\end{figure*}

We then performed the same set of tests for periodic models, with essentially similar results as shown in Figure \ref{fig:periodic_PSD}. The corner plots associated with this set of tests can be found in Figure \ref{fig:periodic_PSD}.

\begin{figure*}
     \centering
        \includegraphics[width=\textwidth]{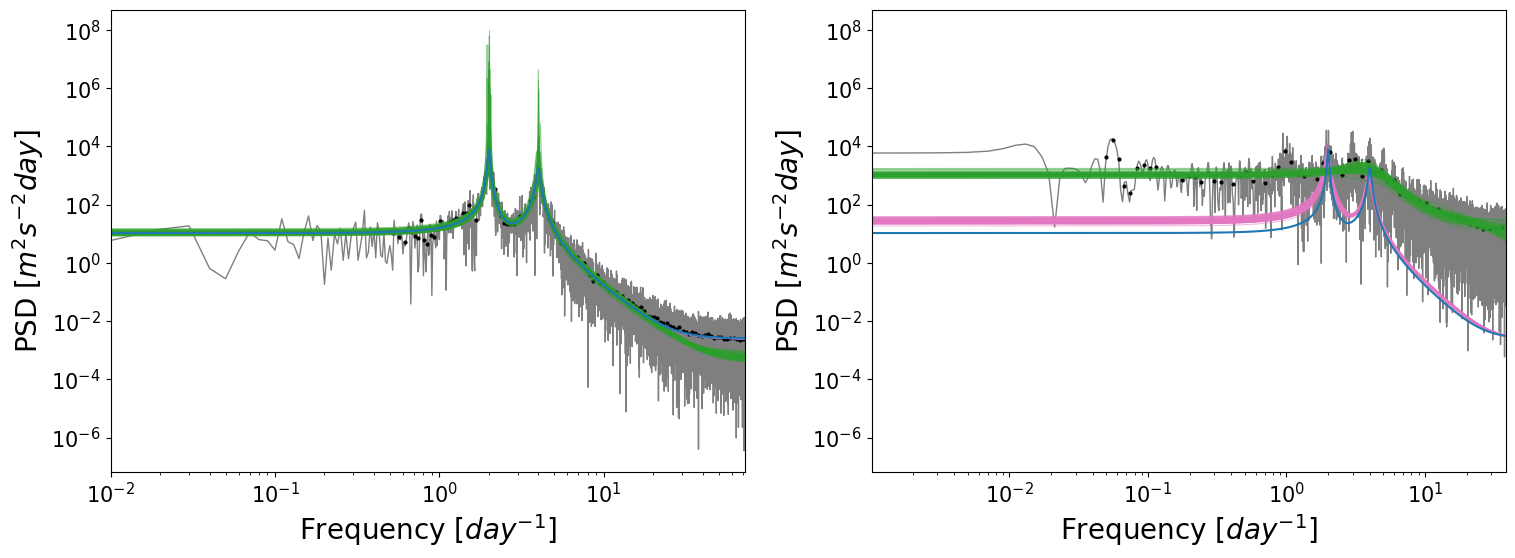}
     \caption{PSD estimates for the \cel\ rotation term model in the regular case (left) and irregular case (right), using the same colour convention as in Figure~\ref{fig:aperiodic_corner}. The injected parameters are as follows : $\sigma =  50$, $P = 0.5$, $Q_0 = 20$, $\delta Q = 15$, $f = 0.8$.  Fitted values can be found in Tables \ref{tab:regular all cases} and \ref{tab:irrregular all cases}. }
     \label{fig:rot_PSD}
\end{figure*}

Finally, we considered a case consisting of two periodic terms, with frequencies differing by a factor 2, as implemented in the \cel\ rotation term mentioned in section~\ref{sec:rotation_term}. The results are shown in Figure~\ref{fig:rot_PSD}.  While both PSD- and GP-fitting were successful in the regularly sampled case (left panel), the PSD fitting approach entirely fails to recover the true shape of the PSD in the irregular case. By contrast, the GP fit still gives tight and accurate constraints on the underlying PSD and its parameters. 

In all cases, the GP-modelling substantially outperformed the PSD-fitting method. Not only do the GP fits result in smaller uncertainties, but they are also substantially more accurate, and generally consistent with the ground truth, unlike the PSD-fitting results. We also note that, as the time-sampling was degraded, MCMC convergence also became increasingly slow in the PSD-fitting case.

\begin{table*}
 	\centering
 	\caption{Injected and fitted parameters for all cases with regular time sampling. Parameters more than 2$\sigma$ away from the true value are highlighted in gray.}
 	\label{tab:regular all cases}
 	\begin{tabular}{c c | c c c c c c} 
 		\hline 
 		\multicolumn{2}{c}{Case}  & \multicolumn{6}{c}{Parameters} \\
 		\hline
       \multirow{4}*{\textbf{Aperiodic}} & \multirow{2}*{Inputted} & $\ln S_0$ & $\ln \nu_0$ &  $\ln \sigma_w$ &  &  & \\
         & & 0.69 & -0.23 & -0.69  &  &  \\ 
        & Fitted PSD & \cellcolor{lightgray}$0.16 \pm 0.12$ & $-0.22 \pm 0.04$ & \cellcolor{lightgray} $-2.15 \pm 0.21$ &  &  &  \\
        & Fitted GP & $0.63 \pm 0.12$  & $-0.22 \pm 0.04 $ &  $-0.69 \pm  - 0.01$ &  &  &  \\
         \hline
        \multirow{4}*{\textbf{Periodic}} & \multirow{2}*{Inputted} & $\ln S_0$ & $\ln \nu_0$ & $\ln Q$ &  $\ln \sigma_w$ &  &  \\
      &  & -2.30 & 1.16 & 2.30 & -0.69 &  &  \\
        & Fitted PSD &$-2.34 \pm 0.14$ & $1.15 \pm 0.02$ &  $2.49^{+0.75}_{-0.44}$ &\cellcolor{lightgray} $-2.65 \pm 0.17$ &  &  \\
        & Fitted GP & $-2.27 \pm 0.04$  & \cellcolor{lightgray} $1.15 \pm 0.00$ &  $2.35 \pm 0.12$ & $-0.69 \pm 0.01$ &  &  \\
         \hline
         \multirow{4}*{\textbf{Rotational}} & \multirow{2}*{Inputted} & $\ln \sigma_0$ & $\ln P$ & $\ln Q_0$ & $\ln dQ$ & $\ln f$ & $\ln \sigma_w$ \\
          & & 3.91 & -0.69 & 3.00 & 2.71 & -0.22 & -0.69  \\
        & Fitted PSD &  $4.55^{+1.97}_{-0.72}$ &  $-0.70 \pm 0.01$ & $2.96^{+0.65}_{-0.48}$ & $5.06^{+4.30}_{-4.97}$ & $-1.48^{+1.35}_{-4.18}$ & \cellcolor{lightgray} $-2.22 \pm 0.25$\\
        & Fitted GP & $3.89 \pm 0.07$ & $-0.69 \pm 0.00$ & $3.03 \pm 0.13$ & $2.22^{+0.58}_{-0.85}$ & $-0.10^{+0.08}_{-0.17}$ & \cellcolor{lightgray}$-0.72 \pm 0.01$\\
         \hline
        \multirow{4}*{\textbf{Periodic + Aperiodic}} & \multirow{2}*{Inputted} & $\ln S_1$ & $\ln \nu_1$ & $\ln Q_1$ & $\ln S_2$ & $\ln \nu_2$ & $\ln \sigma_w$\\
        & &-4.61 & 1.16 & 2.30 & 0 & -0.23 & -0.69\\
        & Fitted PSD & $-4.47 \pm 0.32$ & $1.15 \pm 0.12$ & $2.19^{+0.76}_{-0.46}$ &  $0.13^{+2.73}_{-0.66}$ & $-0.31^{+0.27}_{-0.82}$ &  \cellcolor{lightgray}$-2.69 \pm 0.14$ \\
       &  Fitted GP & $-4.59 \pm 0.08$ & $1.16 \pm 0.01$ & $2.30 \pm 0.12$ & $0.01 \pm 0.13$ & $-0.26 \pm 0.05$ & $-0.69 \pm 0.01$ \\
         \hline
        \multirow{4}*{\textbf{Aperiodic + Aperiodic}} & \multirow{2}*{Inputted} & $\ln S_1$ & $\ln \nu_1$ & $\ln S_2$ & $\ln \nu_2$ & $\ln \sigma_w$ &  \\
        & & 4.61 & 0.87 & 9.21 & -1.14 & -0.69 &  \\
         & Fitted PSD & $4.73 \pm 0.53$ & $0.84 \pm 0.15$ & $9.31 \pm 1.14$ & $-1.31^{+0.38}_{-0.33}$ & \cellcolor{lightgray}$-2.13^{+0.52}_{-0.69}$ &  \\
        & Fitted GP & $4.69 \pm 0.14$ & $0.84 \pm 0.04$ & $9.51 \pm 0.23$ &  $-1.32 \pm 0.09$ & \cellcolor{lightgray}$-0.77 \pm 0.01$ &  \\
         \hline
 	\end{tabular}
 \end{table*}

 \begin{table*}
 	\centering
 	\caption{Injected and fitted parameters for cases with irregular time sampling. Parameters more than 2$\sigma$ away from the true value are highlighted in gray.}
 	\label{tab:irrregular all cases}
 	\begin{tabular}{c c | c c c c c c c} 
 		\hline 
 		\multicolumn{2}{c}{Case}& \multicolumn{6}{|c}{Parameters} \\
 		\hline
 		\multirow{4}*{\textbf{Aperiodic}} & \multirow{2}*{Inputted} & $\ln S_0$ & $\ln \nu_0$ &  $\ln \sigma_w$ &  &  & \\
         & & 0.69 & -0.23 & -0.69  &  &  \\ 
        & Fitted PSD & \cellcolor{lightgray}$2.07 \pm 0.18$ &\cellcolor{lightgray}  $0.72 \pm 0.13$ &\cellcolor{lightgray} $3.51 \pm 0.32$ &  &  &  \\
        & Fitted GP & $0.59 \pm 0.20$ & $-0.21 \pm 0.08$ & $-0.73 \pm 0.02$ &  &  &  \\
         \hline
 		 \multirow{4}*{\textbf{Periodic}} & \multirow{2}*{Inputted} & $\ln S_0$ & $\ln \nu_0$ & $\ln Q$ &  $\ln \sigma_w$ &  &  \\
      &  & -2.30 & 1.16 & 2.30 & -0.69 &  &  \\
        & Fitted PSD &\cellcolor{lightgray}$-0.69 \pm 0.13$ & \cellcolor{lightgray}$1.39 \pm 0.07$ & \cellcolor{lightgray}$0.61 \pm 0.23$ &  \cellcolor{lightgray}$3.47 \pm 0.25$&  & \\
        & Fitted GP & $-2.18 \pm 0.11$ & $1.16 \pm 0.01 $ & $2.01 \pm 0.16$ & $-0.68 \pm 0.02$ &  &   \\
         \hline
         \multirow{4}*{\textbf{Rotational}} & \multirow{2}*{Inputted} & $\ln \sigma_0$ & $\ln P$ & $\ln Q_0$ & $\ln dQ$ & $\ln f$ & $\ln \sigma_w$ \\
         & & 3.91 & -0.69 & 3.00 & 2.71 & -0.22 & -0.69  \\
        & Fitted PSD & \cellcolor{lightgray} $4.49 \pm 0.11$ & \cellcolor{lightgray} $-1.35 \pm 0.10$ & \cellcolor{lightgray} $-5.45^{+1.50}_{-2.57}$ & \cellcolor{lightgray} $-0.47^{+0.41}_{-0.62}$ & $-1.61^{+0.76}_{-0.43}$ &   $0.18^{+3.34}_{-3.41}$\\
        & Fitted GP & $4.00 \pm 0.07$ & $-0.68 \pm 0.01$ &\cellcolor{lightgray}  $2.63 \pm 0.14$ & $-4.79^{+4.26}_{-4.18}$ &  $-0.66 \pm 0.25$ &   \cellcolor{lightgray}$-0.78 \pm 0.04$\\
         \hline
 		\multirow{4}*{\textbf{Periodic + Aperiodic}} & \multirow{2}*{Inputted} & $\ln S_1$ & $\ln \nu_1$ & $\ln Q_1$ & $\ln S_2$ & $\ln \nu_2$ & $\ln \sigma_w$\\
        & &-4.61 & 1.16 & 2.30 & 0 & -0.23 & -0.69\\
        & Fitted PSD & $-4.67^{+6.92}_{-4.47}$ & $1.50 \pm 0.19$ &\cellcolor{lightgray} $-5.17^{+6.51}_{-4.28}$ & \cellcolor{lightgray}$1.10 \pm 0.14$ & \cellcolor{lightgray}$1.22 \pm 0.09$ & \cellcolor{lightgray}$2.55 \pm 0.25$ \\
        & Fitted GP &  $-4.76 \pm 0.24$ & $1.17 \pm 0.01$ & $2.38 \pm 0.25$ & $0.39 \pm 0.26$ & $-0.54 \pm 0.16$ & $-0.69 \pm 0.02$\\
         \hline
 		\multirow{4}*{\textbf{Aperiodic + Aperiodic}} & \multirow{2}*{Inputted} & $\ln S_1$ & $\ln \nu_1$ & $\ln S_2$ & $\ln \nu_2$ & $\ln \sigma_w$ &  \\
        & & 4.61 & 0.87 & 9.21 & -1.14 & -0.69 &  \\
       & Fitted PSD & \cellcolor{lightgray}$8.39^{+0.28}_{-0.57}$ & \cellcolor{lightgray}$1.68^{+0.06}_{-0.10}$ &  $7.65^{+0.98}_{-11.59}$ &  $-0.28^{+0.94}_{-8.33}$ &  $0.69^{+3.43}_{-3.85}$ &  \\
       & Fitted GP & \cellcolor{lightgray} $4.01 \pm 0.29$ & \cellcolor{lightgray}$1.01 \pm 0.08$ & \cellcolor{lightgray} $8.76 \pm 0.27$ & $-1.11 \pm 0.11$ &  $-0.80 \pm 0.06$ &  \\
         \hline
 	\end{tabular}
 \end{table*}

\subsection{Multiple components}
\label{sec: Multiple Components}
We next considered two-component models, to test our ability to disentangle between the components and recover their parameters correctly. We ran two such tests: one with one aperiodic and one periodic component, and one with two aperiodic components. All cases were run multiple times to ensure the repeatability of results. The PSDs plots for these tests can be found in the body of the text in Figure \ref{fig:aperiodic_periodic_PSD}, while the corner plots can be found in Appendix A. It should be noted that in all cases shown below where the MCMC did not converge is due to highly multi-modol posterior distributions.

\begin{figure*}
    \centering
    \includegraphics[width=0.98\textwidth]{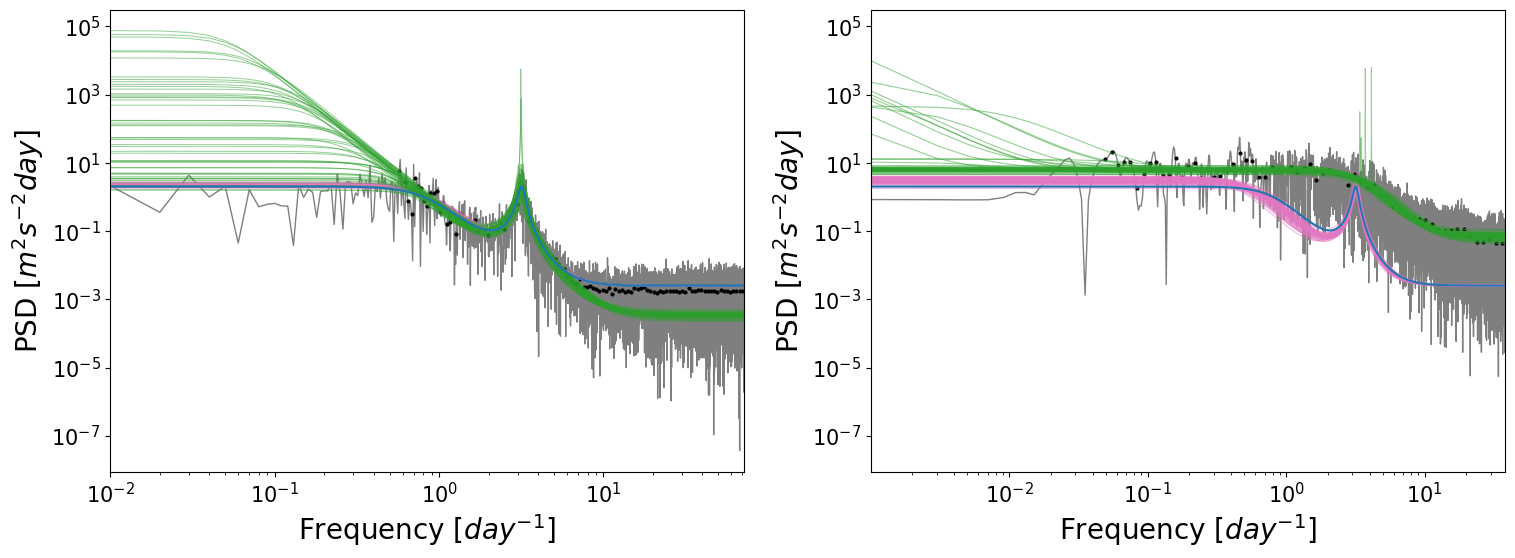}
    \caption{PSD and plots for 2-component models with one periodic and one aperiodic component with regular (left) and irregular (right) sampling,  using the same colour convention as in Figure~\ref{fig:aperiodic_PSD}. The injected parameters are as follows : $S_1 =  0.01$, $\nu_1 = 20/2\pi$, $Q_1 = 10$, $S_2 = 1$, $\nu_2 = 5/2\pi$, $\sigma_w = 0.5$.  Fitted values can be found in Tables \ref{tab:regular all cases} and \ref{tab:irrregular all cases}. }
    \label{fig:aperiodic_periodic_PSD}
\end{figure*}

\begin{figure*}
    \centering
    \includegraphics[width=0.98\textwidth]{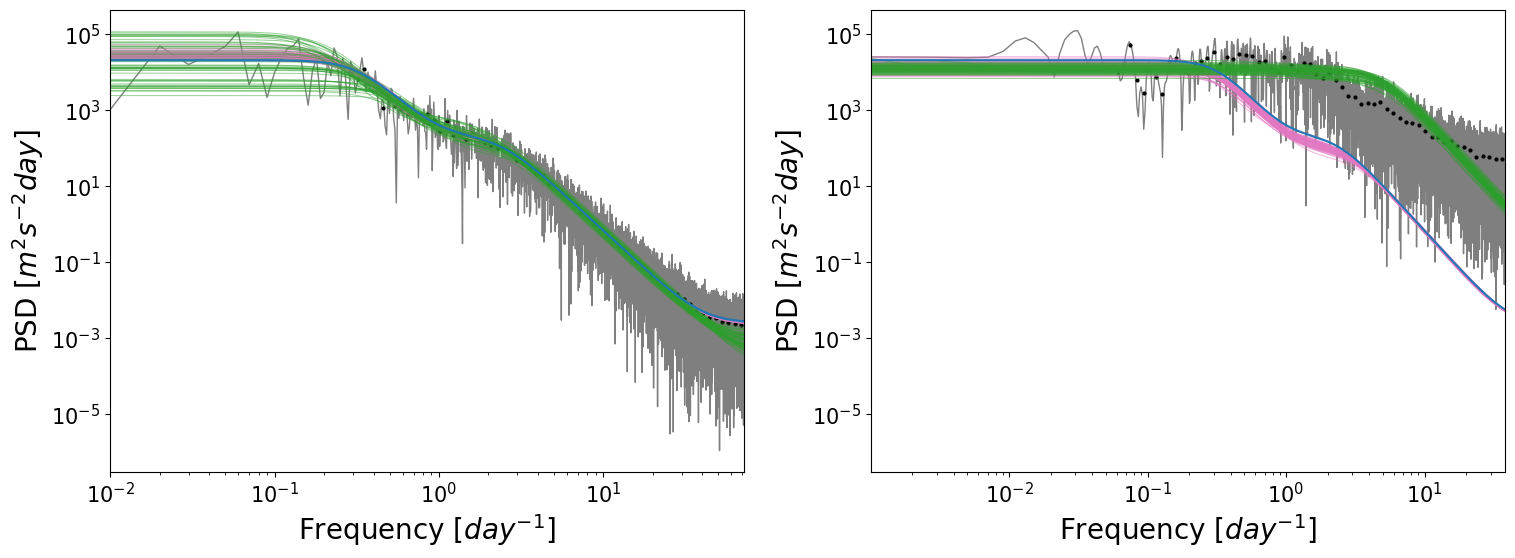}
    \caption{PSD and plots for 2-component models with two aperiodic components with regular (left) and irregular (right) sampling,  using the same colour convention as in Figure~\ref{fig:aperiodic_PSD}. The injected parameters are as follows : $S_1 =  100$, $\nu_1 = 15/2\pi$,  $S_2 = 10,000$, $\nu_2 = 2/2\pi$, $\sigma_w = 0.5$. Fitted values can be found in Tables \ref{tab:regular all cases} and \ref{tab:irrregular all cases}.  }
    \label{fig:aperiodic_aperiodic_PSD}
\end{figure*}

\subsubsection{Model Comparison}

We started by considering a low-frequency, high-amplitude term combined with a higher-frequency, lower amplitude, high-$Q$ periodic term. In the context of asteroseismology, the aperiodic term could represent granulation, and the periodic term the envelope of the $p$-modes. The results of the PSD fit for these tests are shown in Figure ~\ref{fig:aperiodic_periodic_PSD}, and the corresponding corner plots in Appendix \ref{Appendix:Corner Plots}. In both the regular and the irregular time sampling cases, the GP once more outperformed the PSD-fitting method. It is worth noting that, in the regular time sampling case, the PSD-fitting method struggles to constrain the quality factor of the oscillations (this is partly due to the binning of the PSD prior to fitting). Meanwhile, in the irregular case, the oscillations peak is entirely absent from the PSD estimate which naturally leads to poor PSD-fitting results (and the MCMC fails to converge), while the GP fits are much less severely affected: the uncertainties are increased but the results are still consistent with the ground truth. 

An important question is whether the PSD and GP fitting methods can correctly identify the number and nature of the components present in the data. To check whether this is the case we repeated the fits of our 2-component time-series and PSDs using different models, including: (i) 1 periodic component and 1 aperiodic component (true model), (ii) 1 aperiodic component only, (iii) 2 aperiodic components, (iv) 2 periodic components. In each case, we compared the Bayesian Information Criterion (BIC) of the best-fit model. The results of this model comparison are reported in Table~\ref{tab: muliple component BIC}. Both the single aperiodic component model (case ii) and the two aperiodic component model (case iv) where preferred when using PSD-fitting, although the single aperiodic component more so, while the true model (case i) is preferred when using the GP.

\begin{table}
 	\centering
 	\caption{$\Delta$BIC results for multiple component cases in the irregular time sampling regime.  Positive values mean the model is preferred over the true model, negative results mean the opposite. The best fit model is highlighted in gray.}
 	\label{tab: muliple component BIC}
 	\begin{tabular}{ c c  c  c c } 
 		\hline 
 		Case & $\Delta$BIC p + ap & $\Delta$BIC ap & $\Delta$BIC ap + ap & $\Delta$BIC rot \\
 		\hline
 		PSD p + ap & 0 &  \cellcolor{lightgray}40.01 & 8.75 & \\
        GP p + ap &  \cellcolor{lightgray}0 & -45.21 &  -56.50 &\\
         \hline
        PSD ap + ap &  & \cellcolor{lightgray}1.96 & 0 & \\
        GP ap+ ap &  & -3.14 &  \cellcolor{lightgray}0  & \\
        \hline 
        PSD rot &  -48.44 &  \cellcolor{lightgray} 28.92 & -67.91 &  0\\
        GP rot  & -184.53 & -715.01 & -729.66 & \cellcolor{lightgray}0 \\
        \hline 
 	
 	\end{tabular}
 \end{table}

We then considered two aperiodic terms with more similar decay times, representative of super-granulation and granulation. The results of these tests are shown in Figure~\ref{fig:aperiodic_aperiodic_PSD}, with the corner plots found in Appendix \ref{Appendix:Corner Plots}. Here, the "true" parameters are the same order of magnitude as the ones reported by \citet{2023A&A...669A..39A} for granulation and super-granulation in the solar case. It can be seen that the GP fits result in timescales consistent with the ground truth for both granulation and super-granulation, but the PSD fitting approach does not, especially for the granulation-like component. This implies that the timescales reported by A23 should be treated with caution since they were derived using a PSD-fitting procedure very similar to that used in the present work. We also used the BIC to test whether a 2-component model was preferred over a single-component case. The results are reported in Table~\ref{tab: muliple component BIC}. The 1-component model is favoured in the PSD case, however the preference is not statistically significant.

We then proceeded to reduce the frequency separation between the two components, to check when each method stops being able to distinguish between them. The results of these tests are shown in Table \ref{tab: 2 aperiodic degraded sampling}, where the BIC results for models with either 2 aperiodic components or 1 aperiodic component are shown. When using a GP, the true model is preferred, and thus the two components are correctly identified, though the preference over a single component model is statistically significant only down to a frequency ratio of $\frac{\nu_1}{\nu_2} \sim 4$ . By contrast, the PSD fit could only separate the two components for the largest separation. It should be noted that the preference for a 2 aperiodic model in the irregular GP case with $\nu_1 \backslash \nu_2 = 3$ is not statistically significant, and the result may be different depending on the initial time series sample. 

\begin{table}
 	\centering
 	\caption{BIC values for 2 aperiodic components with regular and irregular time sampling as the two components are brought closer together. The only initial parameters that are changed are $\nu_1$ and $\nu_2$. Positive values mean a 1 aperidoic component model is preferred over the true model, these cases are highlighted in gray, negative results mean the opposite. }
 	\label{tab: 2 aperiodic degraded sampling}
 	\begin{tabular}{ c  c c c} 
 		\hline 
 		$\nu_1 \backslash \nu_2$ & fit &  $\Delta$BIC ap \\
        \hline
 		7.5 & PSD reg &   -2.39\\
        7.5 & GP reg &  -234.49 \\
        7.5 & PSD irreg &   -1.54 \\
        7.5 & GP irreg &    -50.79 \\
         \hline
        4.29 & PSD reg &  \cellcolor{lightgray}8.46    \\
        4.29 & GP reg &   -35.29  \\
        4.29 & PSD irreg &  \cellcolor{lightgray}11.28 \\
        4.29 & GP irreg &  -22.06  \\
         \hline
       	3 & PSD reg &  \cellcolor{lightgray}14.32 \\
        3 & GP reg &    \cellcolor{lightgray}11.94 \\
        3 & PSD irreg &  \cellcolor{lightgray}2.02 \\
        3 & GP irreg &  -2.46 \\
        \hline
        2 & PSD reg &   \cellcolor{lightgray}16.95  \\
        2 & GP reg &   \cellcolor{lightgray}17.05   \\
        2 & PSD irreg&   \cellcolor{lightgray}8.26  \\
        2 & GP irreg &    \cellcolor{lightgray}13.88  \\
         \hline
        
 	\end{tabular}
 \end{table}

We also used the BIC test for the signal component  \cel\ rotation term to see if two component models were preferred over the signal component model in this case. The results are reported in Table \ref{tab: muliple component BIC}. A single aperioidc component is preferred in the PSD case. 

\subsubsection{Parameter Accuracy}

Tables \ref{tab:regular all cases} and \ref{tab:irrregular all cases} show the recovered parameters compared to the true values for all 5 cases studied, for both regular and irregular time sampling. Cells for which a derived parameter was more than $2\sigma$ away from the true value are highlighted in gray. In every case discussed above, the GP approach outperformed the PSD-fitting approach. The GP-derived parameters are always close to the true values. However, in a few cases, the uncertainties appear under-estimated, so that the $2\sigma$ credible intervals do not quite overlap with the true value. This tendency to under-estimate uncertainties is particularly pronounced for the white noise parameter and the quality factor of any periodic terms, but is also sometimes observed for the characteristic frequencies. The performance of the PSD-fitting method degrades very noticeable when using irregular sampling, whereas the GP-fitting remains fairly robust, except in the case with two aperiodic components, where both methods failed to recover the granulation frequency and amplitude within $2-\sigma$, though the GP values are still of the correct order of magnitude.

\section{Discussion and conclusions}
\label{sec:discussion}

Having compared the behaviour of PSD-fitting and GP regression on simulated datasets with a known ground truth, we now proceed to discuss the implications of our results, both for RV planet searches (the original motivation for this work) and more widely, and outline perspectives for future work.

\subsection{GP regression versus PSD fitting}

In this work, we have shown that the analytic functions commonly used to model stochastic variability in the light curves and RV observations of the Sun and Sun-like stars can be reproduced, either exactly or approximately, using Gaussian process models. We then showed,q2 using a series of tests on simulated data, that GP regression in the time domain systematically leads to better results than estimating and modelling the PSD in the frequency domain. 

While the GP method leads to more accurate and precise results when applied to regularly sampled data, the difference in performance becomes even more pronounced in the irregular sampling case, where the PSD-fitting results deviate dramatically from the truth, while the GP results are generally consistent with it within $2 \sigma$. The main reason for this is that GP regression does not rely on PSD estimates, which by definition assume a specific model for the signal whose PSD is being estimated. In the case of the GLS periodogram, this model consists of a single sinusoidal signal, plus white noise. When the time sampling is close to regular, individual sines and cosines with frequencies between $1/T$ and $1/2NT$ are approximately mutually orthogonal, and the periodogram closely approximates the true PSD, even if the signal is more complex than a single sinusoid. However, when the sampling becomes sparser, this orthogonality breaks down, and the periodogram is dominated by the window function. This is particularly problematic for stochastic signals, which have a broad intrinsic PSD, and thus deviate strongly from the simple model implicitly assumed by the GLS periodogram. The GP regression approach avoids this problem by obviating the need for a PSD estimate altogether.

It should be noted that the computational cost of GP regression is significantly larger than that of the PSD-fitting approach. For example, for the regularly sampled, single aperiodic component case on a 2020 Mac Mini with an Apple M1 chip, the MCMC sampler completed 580 iterations/second for the PSD-fitting versus 92.5 iterations/second for the GP fitting case, a difference of a factor $>6$. Nonetheless, as the evaluation of the GP likelihood scales linearly with the number of data points for \cel\ models, and the posterior sampling is easy to parallelize, the computational cost is an obstacle that can be overcome, even for fairly large datasets. 

Therefore, we recommend that GP regression in the time domain should be used, wherever possible, in preference to PSD-fitting, to model stochastic processes in astrophysical time-series datasets. One limitation of the GP-fitting method as implemented in this work, which emerged during our tests, is that it has a tendency to under-estimate the uncertainties on the parameters slightly (by a factor of order 2). At this stage, the reason for this tendency is unknown, but we recommend treating these uncertainties with a degree of caution until this behaviour is better understood.

\subsection{Implications for RV planet searches}

The advantages of GP regression over PSD-fitting are particularly relevant to ground-based RV observations, which are always irregularly sampled. Indeed, GP regression is already widely used to model and mitigate activity signals in RV datasets \citep[see e.g.][]{Haywood2014,2015MNRAS.452.2269R,2022MNRAS.509..866B}. The results presented in this paper, and in particular the finding that GPs can robustly identify multi-component signals in irregularly sampled data, indicate that GP regression may also have a role to play in the characterisation and mitigation of other stellar signals such as super-granulation.

Besides the aforementioned GP methods, numerous other approaches to model activity signals at the spectral or line-profile level have been developed in recent years (see e.g.\ \citealt{2021MNRAS.505.1699C,2022A&A...659A..68C,2023A&A...678A...2C,2023arXiv231118326L}). Using a combination of these approaches, it is widely expected that future Extreme Precision RV (EPRV) surveys aiming to detect Earth analogues, such as the Terra Hunting Experiment (THE, \citealt{2018MNRAS.479.2968H}), which will benefit from unprecedented time-sampling and high signal-to-noise ratio, will succeed in mitigating activity signals in Sun-like stars from several m/s down to the sub-m/s level. 

The next major challenge is thus super-granulation \citep{2020A&A...642A.157M} for which no effective correction methods have been developed yet. Indeed, \citet{2024MNRAS.527.7681L} recently analysed the RV variability of the quiet Sun as observed by HARPS-N (after accounting for the effects of active regions using SDO data, following the method developed by \citealt{2016MNRAS.457.3637H} and \citealt{2019ApJ...874..107M}). During the low-activity part of the Solar cycle, they found that the quiet photosphere contribution to RV variability is as important as, if not larger than, that from active regions.

A natural continuation of this work will thus be to use the GP methodology we have proposed here to revisit recent studies aiming to quantify the timescales and amplitudes of variability due to (super-)granulation on the Sun, including A23 and \citet{2024MNRAS.527.7681L}, as well as in other stars (see for example \citealt{2023A&A...670A..24S}, who compared granulation signals in photometry and RV using simultaneous observations with CHEOPS and VLT/ESPRESSO). While A23 and \citet{2023A&A...670A..24S} used PSD-fitting, \citet{2024MNRAS.527.7681L} used structure functions, a non-parametric approach that, like GP regression, can be applied directly to irregularly sampled data, but doesn't explicitly use a generative model, and it would be valuable to gain a better understanding of the relationship between structure functions and GP models. One potentially interesting avenue for future progress may also be to search for and characterise any non-Gaussianity in stellar RV signals.

\subsection{Asteroseismology and planetary transits}

The PSD models discussed in Section~\ref{sec:psd_review} first arose in the context of helioseismology, and are widely used in asteroseismology, whether using RV observations or photometry. While the "background" in this context is merely a nuisance signal to the oscillations, it is conceivable that there may be some advantage to be gained by modelling this background directly in the time domain using GPs rather than by evaluating and fitting a PSD. This may be particularly helpful for later-type main sequence stars, for which the oscillation frequencies are higher and the amplitudes lower than earlier type or evolved stars \citep[see e.g.][]{2013ARA&A..51..353C}. As we have shown in this work, multi-component GP models can be used to detect (quasi-)periodic signals in the presence of correlated noise, even when no clear peak is visible in the PSD of the data. Such models, consisting of one or more aperiodic terms to represent (super-)granulation and a single high quality-factor term to represent the envelope of the $p$-modes, may help push the detection threshold for $p$-modes beyond early-$K$ spectral types, which is the current limit.

This type of model has already been used to model granulation signals and oscillations simultaneously in order to characterise planetary transits around red giants \citep{2017AJ....154..254G}, and may likely prove important for precise measurements of planetary radii from transits \citep{2020A&A...634A..75B}, e.g. in the context of the PLATO space mission \citep{2016AN....337..961R}.

\subsubsection{Quasi-Periodic Oscillations and accreting systems}

Accreting systems on all scales, from compact binaries to Active Galactic Nuclei (AGN) display Quasi-Periodic Oscillations (QPOs) \citep[see e.g.][]{2011Ap&SS.332....1R,2019NewAR..8501524I}. QPOs are time-variable quasi-periodic signals observed at X-ray, UV and optical wavelengths, and are usually modelled in the frequency domain as a sum of Lorentzian functions \citep[][]{2002ApJ...572..392B, 2002ApJ...568..912V}. The parameters of these models, including the frequency and amplitude of any quasi-periodic components as well as the frequency at which aperiodic components display breaks in their PSDs, are used to interpret QPOs and understand their origin and diversity. There is every reason to expect that standard QPO modelling procedures are subject to the same limitations as outlined in this work for stellar signals in RV or photometry, and that GP regression could be a more robust alternative, particularly for shorter or more sparsely sampled time-series. Indeed, GP regression has begun to be used more frequently in the context of QPOs and AGN variability \citep[see][and references therein]{2023ARA&A..61..329A}. 

\section*{Acknowledgements}

This publication is part of a project that has received funding from the European Research Council (ERC) under the European Union’s Horizon 2020 research and innovation program (Grant agreement No. 865624). NKOS thanks the LSSTC Data Science Fellowship Program, which is funded by LSSTC, NSF Cybertraining Grant number 1829740, the Brinson Foundation, and the Moore Foundation; her participation in the program has benefited this work. This work made use of \texttt{numpy} \citep[][]{numpy}, \texttt{matplotlib} \citep[][]{matplotlib}, and \texttt{pandas} \citep{pandas} libraries. This work made use of Astropy:\footnote{\url{http://www.astropy.org}} a community-developed core Python package and an ecosystem of tools and resources for astronomy \citep{astropy1, astropy2,astropy3}.

\section*{Data Availability}
 
The codes used in this manuscript are readily available at  \href{https://github.com/NiamhOSullivan}{https://github.com/NiamhOSullivan}. 



\bibliographystyle{mnras}
\bibliography{main_text} 

\begin{thebibliography}{}
\makeatletter
\relax
\def\mn@urlcharsother{\let\do\@makeother \do\$\do\&\do\#\do\^\do\_\do\%\do\~}
\def\mn@doi{\begingroup\mn@urlcharsother \@ifnextchar [ {\mn@doi@} {\mn@doi@[]}}
\def\mn@doi@[#1]#2{\def\@tempa{#1}\ifx\@tempa\@empty \href {http://dx.doi.org/#2} {doi:#2}\else \href {http://dx.doi.org/#2} {#1}\fi \endgroup}
\def\mn@eprint#1#2{\mn@eprint@#1:#2::\@nil}
\def\mn@eprint@arXiv#1{\href {http://arxiv.org/abs/#1} {{\tt arXiv:#1}}}
\def\mn@eprint@dblp#1{\href {http://dblp.uni-trier.de/rec/bibtex/#1.xml} {dblp:#1}}
\def\mn@eprint@#1:#2:#3:#4\@nil{\def\@tempa {#1}\def\@tempb {#2}\def\@tempc {#3}\ifx \@tempc \@empty \let \@tempc \@tempb \let \@tempb \@tempa \fi \ifx \@tempb \@empty \def\@tempb {arXiv}\fi \@ifundefined {mn@eprint@\@tempb}{\@tempb:\@tempc}{\expandafter \expandafter \csname mn@eprint@\@tempb\endcsname \expandafter{\@tempc}}}

\bibitem[\protect\citeauthoryear{{Aigrain} \& {Foreman-Mackey}}{{Aigrain} \& {Foreman-Mackey}}{2023}]{2023ARA&A..61..329A}
{Aigrain} S.,  {Foreman-Mackey} D.,  2023, \mn@doi [\araa] {10.1146/annurev-astro-052920-103508}, \href {https://ui.adsabs.harvard.edu/abs/2023ARA&A..61..329A} {61, 329}

\bibitem[\protect\citeauthoryear{{Aigrain}, {Favata}  \& {Gilmore}}{{Aigrain} et~al.}{2004}]{2004A&A...414.1139A}
{Aigrain} S.,  {Favata} F.,   {Gilmore} G.,  2004, \mn@doi [\aap] {10.1051/0004-6361:20034039}, \href {https://ui.adsabs.harvard.edu/abs/2004A&A...414.1139A} {414, 1139}

\bibitem[\protect\citeauthoryear{{Al Moulla}, {Dumusque}, {Figueira}, {Lo Curto}, {Santos}  \& {Wildi}}{{Al Moulla} et~al.}{2023}]{2023A&A...669A..39A}
{Al Moulla} K.,  {Dumusque} X.,  {Figueira} P.,  {Lo Curto} G.,  {Santos} N.~C.,   {Wildi} F.,  2023, \mn@doi [\aap] {10.1051/0004-6361/202244663}, \href {https://ui.adsabs.harvard.edu/abs/2023A&A...669A..39A} {669, A39}

\bibitem[\protect\citeauthoryear{{Astropy Collaboration} et~al.,}{{Astropy Collaboration} et~al.}{2013}]{astropy1}
{Astropy Collaboration} et~al., 2013, \mn@doi [\aap] {10.1051/0004-6361/201322068}, \href {https://ui.adsabs.harvard.edu/abs/2013A&A...558A..33A} {558, A33}

\bibitem[\protect\citeauthoryear{{Astropy Collaboration} et~al.,}{{Astropy Collaboration} et~al.}{2018}]{astropy2}
{Astropy Collaboration} et~al., 2018, \mn@doi [\aj] {10.3847/1538-3881/aabc4f}, \href {https://ui.adsabs.harvard.edu/abs/2018AJ....156..123A} {156, 123}

\bibitem[\protect\citeauthoryear{{Astropy Collaboration} et~al.,}{{Astropy Collaboration} et~al.}{2022}]{astropy3}
{Astropy Collaboration} et~al., 2022, \mn@doi [\apj] {10.3847/1538-4357/ac7c74}, \href {https://ui.adsabs.harvard.edu/abs/2022ApJ...935..167A} {935, 167}

\bibitem[\protect\citeauthoryear{{Barrag{\'a}n}, {Aigrain}, {Rajpaul}  \& {Zicher}}{{Barrag{\'a}n} et~al.}{2022}]{2022MNRAS.509..866B}
{Barrag{\'a}n} O.,  {Aigrain} S.,  {Rajpaul} V.~M.,   {Zicher} N.,  2022, \mn@doi [\mnras] {10.1093/mnras/stab2889}, \href {https://ui.adsabs.harvard.edu/abs/2022MNRAS.509..866B} {509, 866}

\bibitem[\protect\citeauthoryear{{Barros}, {Demangeon}, {D{\'\i}az}, {Cabrera}, {Santos}, {Faria}  \& {Pereira}}{{Barros} et~al.}{2020}]{2020A&A...634A..75B}
{Barros} S.~C.~C.,  {Demangeon} O.,  {D{\'\i}az} R.~F.,  {Cabrera} J.,  {Santos} N.~C.,  {Faria} J.~P.,   {Pereira} F.,  2020, \mn@doi [\aap] {10.1051/0004-6361/201936086}, \href {https://ui.adsabs.harvard.edu/abs/2020A&A...634A..75B} {634, A75}

\bibitem[\protect\citeauthoryear{{Belloni}, {Psaltis}  \& {van der Klis}}{{Belloni} et~al.}{2002}]{2002ApJ...572..392B}
{Belloni} T.,  {Psaltis} D.,   {van der Klis} M.,  2002, \mn@doi [\apj] {10.1086/340290}, \href {https://ui.adsabs.harvard.edu/abs/2002ApJ...572..392B} {572, 392}

\bibitem[\protect\citeauthoryear{{Chaplin} \& {Miglio}}{{Chaplin} \& {Miglio}}{2013}]{2013ARA&A..51..353C}
{Chaplin} W.~J.,  {Miglio} A.,  2013, \mn@doi [\araa] {10.1146/annurev-astro-082812-140938}, \href {https://ui.adsabs.harvard.edu/abs/2013ARA&A..51..353C} {51, 353}

\bibitem[\protect\citeauthoryear{{Collier Cameron} et~al.,}{{Collier Cameron} et~al.}{2021}]{2021MNRAS.505.1699C}
{Collier Cameron} A.,  et~al., 2021, \mn@doi [\mnras] {10.1093/mnras/stab1323}, \href {https://ui.adsabs.harvard.edu/abs/2021MNRAS.505.1699C} {505, 1699}

\bibitem[\protect\citeauthoryear{{Cretignier}, {Dumusque}  \& {Pepe}}{{Cretignier} et~al.}{2022}]{2022A&A...659A..68C}
{Cretignier} M.,  {Dumusque} X.,   {Pepe} F.,  2022, \mn@doi [\aap] {10.1051/0004-6361/202142435}, \href {https://ui.adsabs.harvard.edu/abs/2022A&A...659A..68C} {659, A68}

\bibitem[\protect\citeauthoryear{{Cretignier}, {Dumusque}, {Aigrain}  \& {Pepe}}{{Cretignier} et~al.}{2023}]{2023A&A...678A...2C}
{Cretignier} M.,  {Dumusque} X.,  {Aigrain} S.,   {Pepe} F.,  2023, \mn@doi [\aap] {10.1051/0004-6361/202347232}, \href {https://ui.adsabs.harvard.edu/abs/2023A&A...678A...2C} {678, A2}

\bibitem[\protect\citeauthoryear{{Dumusque}, {Udry}, {Lovis}, {Santos}  \& {Monteiro}}{{Dumusque} et~al.}{2011}]{2011A&A...525A.140D}
{Dumusque} X.,  {Udry} S.,  {Lovis} C.,  {Santos} N.~C.,   {Monteiro} M.~J.~P.~F.~G.,  2011, \mn@doi [\aap] {10.1051/0004-6361/201014097}, \href {https://ui.adsabs.harvard.edu/abs/2011A&A...525A.140D} {525, A140}

\bibitem[\protect\citeauthoryear{{Dumusque} et~al.,}{{Dumusque} et~al.}{2021}]{2021A&A...648A.103D}
{Dumusque} X.,  et~al., 2021, \mn@doi [\aap] {10.1051/0004-6361/202039350}, \href {https://ui.adsabs.harvard.edu/abs/2021A&A...648A.103D} {648, A103}

\bibitem[\protect\citeauthoryear{Foreman-Mackey}{Foreman-Mackey}{2016}]{corner}
Foreman-Mackey D.,  2016, \mn@doi [The Journal of Open Source Software] {10.21105/joss.00024}, 1, 24

\bibitem[\protect\citeauthoryear{{Foreman-Mackey}}{{Foreman-Mackey}}{2018}]{celerite2}
{Foreman-Mackey} D.,  2018, \mn@doi [Research Notes of the American Astronomical Society] {10.3847/2515-5172/aaaf6c}, \href {http://adsabs.harvard.edu/abs/2018RNAAS...2a..31F} {2, 31}

\bibitem[\protect\citeauthoryear{{Foreman-Mackey}, {Hogg}, {Lang}  \& {Goodman}}{{Foreman-Mackey} et~al.}{2013}]{2013PASP..125..306F}
{Foreman-Mackey} D.,  {Hogg} D.~W.,  {Lang} D.,   {Goodman} J.,  2013, \mn@doi [\pasp] {10.1086/670067}, \href {https://ui.adsabs.harvard.edu/abs/2013PASP..125..306F} {125, 306}

\bibitem[\protect\citeauthoryear{{Foreman-Mackey}, {Agol}, {Ambikasaran}  \& {Angus}}{{Foreman-Mackey} et~al.}{2017}]{celerite1}
{Foreman-Mackey} D.,  {Agol} E.,  {Ambikasaran} S.,   {Angus} R.,  2017, \mn@doi [\aj] {10.3847/1538-3881/aa9332}, \href {http://adsabs.harvard.edu/abs/2017AJ....154..220F} {154, 220}

\bibitem[\protect\citeauthoryear{{Foreman-Mackey} et~al.,}{{Foreman-Mackey} et~al.}{2019}]{2019JOSS....4.1864F}
{Foreman-Mackey} D.,  et~al., 2019, \mn@doi [The Journal of Open Source Software] {10.21105/joss.01864}, \href {https://ui.adsabs.harvard.edu/abs/2019JOSS....4.1864F} {4, 1864}

\bibitem[\protect\citeauthoryear{Foreman-Mackey et~al.,}{Foreman-Mackey et~al.}{2021}]{Foreman-Mackey2021}
Foreman-Mackey D.,  et~al., 2021, \mn@doi [Journal of Open Source Software] {10.21105/joss.03285}, 6, 3285

\bibitem[\protect\citeauthoryear{{Goldreich}, {Murray}  \& {Kumar}}{{Goldreich} et~al.}{1994}]{1994ApJ...424..466G}
{Goldreich} P.,  {Murray} N.,   {Kumar} P.,  1994, \mn@doi [\apj] {10.1086/173904}, \href {https://ui.adsabs.harvard.edu/abs/1994ApJ...424..466G} {424, 466}

\bibitem[\protect\citeauthoryear{{Grunblatt} et~al.,}{{Grunblatt} et~al.}{2017}]{2017AJ....154..254G}
{Grunblatt} S.~K.,  et~al., 2017, \mn@doi [\aj] {10.3847/1538-3881/aa932d}, \href {https://ui.adsabs.harvard.edu/abs/2017AJ....154..254G} {154, 254}

\bibitem[\protect\citeauthoryear{{Hall}, {Thompson}, {Handley}  \& {Queloz}}{{Hall} et~al.}{2018}]{2018MNRAS.479.2968H}
{Hall} R.~D.,  {Thompson} S.~J.,  {Handley} W.,   {Queloz} D.,  2018, \mn@doi [\mnras] {10.1093/mnras/sty1464}, \href {https://ui.adsabs.harvard.edu/abs/2018MNRAS.479.2968H} {479, 2968}

\bibitem[\protect\citeauthoryear{Harris et~al.,}{Harris et~al.}{2020}]{numpy}
Harris C.~R.,  et~al., 2020, \mn@doi [Nature] {10.1038/s41586-020-2649-2}, 585, 357

\bibitem[\protect\citeauthoryear{{Harvey}}{{Harvey}}{1985}]{1985ESASP.235..199H}
{Harvey} J.,  1985, in {Rolfe} E.,  {Battrick} B.,  eds,  ESA Special Publication Vol. 235, Future Missions in Solar, Heliospheric \& Space Plasma Physics. p.~199

\bibitem[\protect\citeauthoryear{{Harvey}, {Duvall}, {Jefferies}  \& {Pomerantz}}{{Harvey} et~al.}{1993}]{1993ASPC...42..111H}
{Harvey} J.~W.,  {Duvall} T.~L. J.,  {Jefferies} S.~M.,   {Pomerantz} M.~A.,  1993, in {Brown} T.~M.,  ed.,  Astronomical Society of the Pacific Conference Series Vol. 42, GONG 1992. Seismic Investigation of the Sun and Stars. p.~111

\bibitem[\protect\citeauthoryear{{Haywood} et~al.,}{{Haywood} et~al.}{2014}]{Haywood2014}
{Haywood} R.~D.,  et~al., 2014, \mn@doi [\mnras] {10.1093/mnras/stu1320}, \href {https://ui.adsabs.harvard.edu/abs/2014MNRAS.443.2517H} {443, 2517}

\bibitem[\protect\citeauthoryear{{Haywood} et~al.,}{{Haywood} et~al.}{2016}]{2016MNRAS.457.3637H}
{Haywood} R.~D.,  et~al., 2016, \mn@doi [\mnras] {10.1093/mnras/stw187}, \href {https://ui.adsabs.harvard.edu/abs/2016MNRAS.457.3637H} {457, 3637}

\bibitem[\protect\citeauthoryear{Hunter}{Hunter}{2007}]{matplotlib}
Hunter J.~D.,  2007, \mn@doi [Computing in Science & Engineering] {10.1109/MCSE.2007.55}, 9, 90

\bibitem[\protect\citeauthoryear{{Ingram} \& {Motta}}{{Ingram} \& {Motta}}{2019}]{2019NewAR..8501524I}
{Ingram} A.~R.,  {Motta} S.~E.,  2019, \mn@doi [\nar] {10.1016/j.newar.2020.101524}, \href {https://ui.adsabs.harvard.edu/abs/2019NewAR..8501524I} {85, 101524}

\bibitem[\protect\citeauthoryear{{Kallinger} et~al.,}{{Kallinger} et~al.}{2014}]{2014A&A...570A..41K}
{Kallinger} T.,  et~al., 2014, \mn@doi [\aap] {10.1051/0004-6361/201424313}, \href {https://ui.adsabs.harvard.edu/abs/2014A&A...570A..41K} {570, A41}

\bibitem[\protect\citeauthoryear{{Karoff} et~al.,}{{Karoff} et~al.}{2013}]{2013ApJ...767...34K}
{Karoff} C.,  et~al., 2013, \mn@doi [\apj] {10.1088/0004-637X/767/1/34}, \href {https://ui.adsabs.harvard.edu/abs/2013ApJ...767...34K} {767, 34}

\bibitem[\protect\citeauthoryear{{Kjeldsen} \& {Bedding}}{{Kjeldsen} \& {Bedding}}{1995}]{1995A&A...293...87K}
{Kjeldsen} H.,  {Bedding} T.~R.,  1995, \mn@doi [\aap] {10.48550/arXiv.astro-ph/9403015}, \href {https://ui.adsabs.harvard.edu/abs/1995A&A...293...87K} {293, 87}

\bibitem[\protect\citeauthoryear{{Lakeland} et~al.,}{{Lakeland} et~al.}{2024}]{2024MNRAS.527.7681L}
{Lakeland} B.~S.,  et~al., 2024, \mn@doi [\mnras] {10.1093/mnras/stad3723}, \href {https://ui.adsabs.harvard.edu/abs/2024MNRAS.527.7681L} {527, 7681}

\bibitem[\protect\citeauthoryear{{Lefebvre}, {Garc{\'\i}a}, {Jim{\'e}nez-Reyes}, {Turck-Chi{\`e}ze}  \& {Mathur}}{{Lefebvre} et~al.}{2008}]{2008A&A...490.1143L}
{Lefebvre} S.,  {Garc{\'\i}a} R.~A.,  {Jim{\'e}nez-Reyes} S.~J.,  {Turck-Chi{\`e}ze} S.,   {Mathur} S.,  2008, \mn@doi [\aap] {10.1051/0004-6361:200810344}, \href {https://ui.adsabs.harvard.edu/abs/2008A&A...490.1143L} {490, 1143}

\bibitem[\protect\citeauthoryear{{Liang}, {Winn}  \& {Melchior}}{{Liang} et~al.}{2023}]{2023arXiv231118326L}
{Liang} Y.,  {Winn} J.~N.,   {Melchior} P.,  2023, \mn@doi [arXiv e-prints] {10.48550/arXiv.2311.18326}, \href {https://ui.adsabs.harvard.edu/abs/2023arXiv231118326L} {p. arXiv:2311.18326}

\bibitem[\protect\citeauthoryear{{Lomb}}{{Lomb}}{1976}]{1976Ap&SS..39..447L}
{Lomb} N.~R.,  1976, \mn@doi [\apss] {10.1007/BF00648343}, \href {https://ui.adsabs.harvard.edu/abs/1976Ap&SS..39..447L} {39, 447}

\bibitem[\protect\citeauthoryear{{Meunier}}{{Meunier}}{2021}]{2021arXiv210406072M}
{Meunier} N.,  2021, \mn@doi [arXiv e-prints] {10.48550/arXiv.2104.06072}, \href {https://ui.adsabs.harvard.edu/abs/2021arXiv210406072M} {p. arXiv:2104.06072}

\bibitem[\protect\citeauthoryear{{Meunier} \& {Lagrange}}{{Meunier} \& {Lagrange}}{2020}]{2020A&A...642A.157M}
{Meunier} N.,  {Lagrange} A.~M.,  2020, \mn@doi [\aap] {10.1051/0004-6361/202038376}, \href {https://ui.adsabs.harvard.edu/abs/2020A&A...642A.157M} {642, A157}

\bibitem[\protect\citeauthoryear{{Meunier}, {Desort}  \& {Lagrange}}{{Meunier} et~al.}{2010}]{2010A&A...512A..39M}
{Meunier} N.,  {Desort} M.,   {Lagrange} A.~M.,  2010, \mn@doi [\aap] {10.1051/0004-6361/200913551}, \href {https://ui.adsabs.harvard.edu/abs/2010A&A...512A..39M} {512, A39}

\bibitem[\protect\citeauthoryear{{Meunier}, {Lagrange}, {Borgniet}  \& {Rieutord}}{{Meunier} et~al.}{2015}]{2015A&A...583A.118M}
{Meunier} N.,  {Lagrange} A.~M.,  {Borgniet} S.,   {Rieutord} M.,  2015, \mn@doi [\aap] {10.1051/0004-6361/201525721}, \href {https://ui.adsabs.harvard.edu/abs/2015A&A...583A.118M} {583, A118}

\bibitem[\protect\citeauthoryear{{Michel}, {Samadi}, {Baudin}, {Barban}, {Appourchaux}  \& {Auvergne}}{{Michel} et~al.}{2009}]{2009A&A...495..979M}
{Michel} E.,  {Samadi} R.,  {Baudin} F.,  {Barban} C.,  {Appourchaux} T.,   {Auvergne} M.,  2009, \mn@doi [\aap] {10.1051/0004-6361:200810353}, \href {https://ui.adsabs.harvard.edu/abs/2009A&A...495..979M} {495, 979}

\bibitem[\protect\citeauthoryear{{Milbourne} et~al.,}{{Milbourne} et~al.}{2019}]{2019ApJ...874..107M}
{Milbourne} T.~W.,  et~al., 2019, \mn@doi [\apj] {10.3847/1538-4357/ab064a}, \href {https://ui.adsabs.harvard.edu/abs/2019ApJ...874..107M} {874, 107}

\bibitem[\protect\citeauthoryear{{Nordlund}, {Spruit}, {Ludwig}  \& {Trampedach}}{{Nordlund} et~al.}{1997}]{1997A&A...328..229N}
{Nordlund} A.,  {Spruit} H.~C.,  {Ludwig} H.~G.,   {Trampedach} R.,  1997, \aap, \href {https://ui.adsabs.harvard.edu/abs/1997A&A...328..229N} {328, 229}

\bibitem[\protect\citeauthoryear{{Rajpaul}, {Aigrain}, {Osborne}, {Reece}  \& {Roberts}}{{Rajpaul} et~al.}{2015}]{2015MNRAS.452.2269R}
{Rajpaul} V.,  {Aigrain} S.,  {Osborne} M.~A.,  {Reece} S.,   {Roberts} S.,  2015, \mn@doi [\mnras] {10.1093/mnras/stv1428}, \href {https://ui.adsabs.harvard.edu/abs/2015MNRAS.452.2269R} {452, 2269}

\bibitem[\protect\citeauthoryear{{Rauer}, {Aerts}, {Cabrera}  \& {PLATO Team}}{{Rauer} et~al.}{2016}]{2016AN....337..961R}
{Rauer} H.,  {Aerts} C.,  {Cabrera} J.,   {PLATO Team} 2016, \mn@doi [Astronomische Nachrichten] {10.1002/asna.201612408}, \href {https://ui.adsabs.harvard.edu/abs/2016AN....337..961R} {337, 961}

\bibitem[\protect\citeauthoryear{{Reig}}{{Reig}}{2011}]{2011Ap&SS.332....1R}
{Reig} P.,  2011, \mn@doi [\apss] {10.1007/s10509-010-0575-8}, \href {https://ui.adsabs.harvard.edu/abs/2011Ap&SS.332....1R} {332, 1}

\bibitem[\protect\citeauthoryear{{Rieutord} \& {Rincon}}{{Rieutord} \& {Rincon}}{2010}]{2010LRSP....7....2R}
{Rieutord} M.,  {Rincon} F.,  2010, \mn@doi [Living Reviews in Solar Physics] {10.12942/lrsp-2010-2}, \href {https://ui.adsabs.harvard.edu/abs/2010LRSP....7....2R} {7, 2}

\bibitem[\protect\citeauthoryear{{Rimmele}, {Goode}, {Harold}  \& {Stebbins}}{{Rimmele} et~al.}{1995}]{1995ApJ...444L.119R}
{Rimmele} T.~R.,  {Goode} P.~R.,  {Harold} E.,   {Stebbins} R.~T.,  1995, \mn@doi [\apjl] {10.1086/187874}, \href {https://ui.adsabs.harvard.edu/abs/1995ApJ...444L.119R} {444, L119}

\bibitem[\protect\citeauthoryear{{Roudier}, {Vigneau}, {Espagnet}, {Muller}, {Mein}  \& {Malherbe}}{{Roudier} et~al.}{1991}]{1991A&A...248..245R}
{Roudier} T.,  {Vigneau} J.,  {Espagnet} O.,  {Muller} R.,  {Mein} P.,   {Malherbe} J.~M.,  1991, \aap, \href {https://ui.adsabs.harvard.edu/abs/1991A&A...248..245R} {248, 245}

\bibitem[\protect\citeauthoryear{{Saar} \& {Donahue}}{{Saar} \& {Donahue}}{1997}]{1997ApJ...485..319S}
{Saar} S.~H.,  {Donahue} R.~A.,  1997, \mn@doi [\apj] {10.1086/304392}, \href {https://ui.adsabs.harvard.edu/abs/1997ApJ...485..319S} {485, 319}

\bibitem[\protect\citeauthoryear{{Santos}, {Gomes da Silva}, {Lovis}  \& {Melo}}{{Santos} et~al.}{2010}]{2010A&A...511A..54S}
{Santos} N.~C.,  {Gomes da Silva} J.,  {Lovis} C.,   {Melo} C.,  2010, \mn@doi [\aap] {10.1051/0004-6361/200913433}, \href {https://ui.adsabs.harvard.edu/abs/2010A&A...511A..54S} {511, A54}

\bibitem[\protect\citeauthoryear{{Scargle}}{{Scargle}}{1982}]{1982ApJ...263..835S}
{Scargle} J.~D.,  1982, \mn@doi [\apj] {10.1086/160554}, \href {https://ui.adsabs.harvard.edu/abs/1982ApJ...263..835S} {263, 835}

\bibitem[\protect\citeauthoryear{{Sulis} et~al.,}{{Sulis} et~al.}{2023}]{2023A&A...670A..24S}
{Sulis} S.,  et~al., 2023, \mn@doi [\aap] {10.1051/0004-6361/202244223}, \href {https://ui.adsabs.harvard.edu/abs/2023A&A...670A..24S} {670, A24}

\bibitem[\protect\citeauthoryear{{VanderPlas}}{{VanderPlas}}{2018}]{2018ApJS..236...16V}
{VanderPlas} J.~T.,  2018, \mn@doi [\apjs] {10.3847/1538-4365/aab766}, \href {https://ui.adsabs.harvard.edu/abs/2018ApJS..236...16V} {236, 16}

\bibitem[\protect\citeauthoryear{{Zechmeister} \& {K{\"u}rster}}{{Zechmeister} \& {K{\"u}rster}}{2009}]{2009A&A...496..577Z}
{Zechmeister} M.,  {K{\"u}rster} M.,  2009, \mn@doi [\aap] {10.1051/0004-6361:200811296}, \href {https://ui.adsabs.harvard.edu/abs/2009A&A...496..577Z} {496, 577}

\bibitem[\protect\citeauthoryear{pandas~development team}{pandas~development team}{2020}]{pandas}
pandas~development team T.,  2020, pandas-dev/pandas: Pandas, \mn@doi{10.5281/zenodo.3509134}, \url {https://doi.org/10.5281/zenodo.3509134}

\bibitem[\protect\citeauthoryear{{van Straaten}, {van der Klis}, {di Salvo}  \& {Belloni}}{{van Straaten} et~al.}{2002}]{2002ApJ...568..912V}
{van Straaten} S.,  {van der Klis} M.,  {di Salvo} T.,   {Belloni} T.,  2002, \mn@doi [\apj] {10.1086/338948}, \href {https://ui.adsabs.harvard.edu/abs/2002ApJ...568..912V} {568, 912}

\makeatother
\end{thebibliography}


\appendix
\section{Corner Plots}
\label{Appendix:Corner Plots}

Corner plots of the parameters of the PSDs shown in Figures \ref{fig:periodic_PSD}, \ref{fig:rot_PSD}, \ref{fig:aperiodic_periodic_PSD}, and \ref{fig:aperiodic_aperiodic_PSD}.

\begin{figure*}
    \centering
    \includegraphics[width=0.98\textwidth]{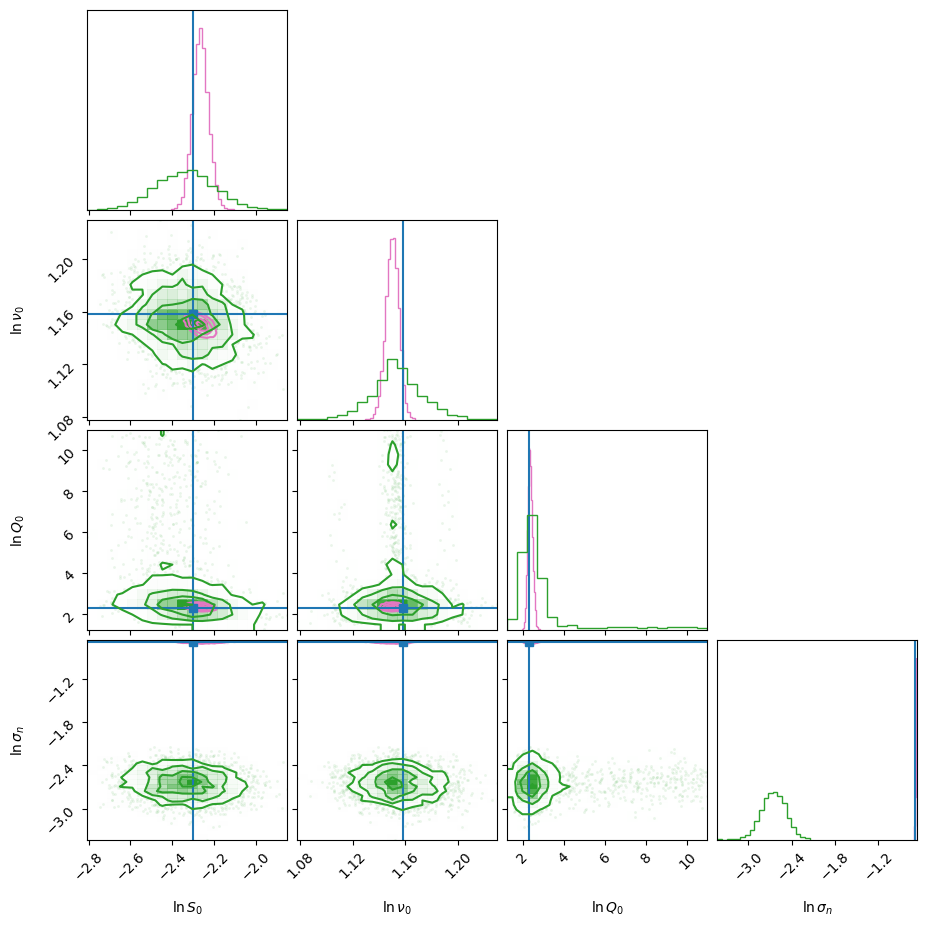}
    \caption{Corner plots for the case of a single, periodic component with $Q=10$ $S_0 = 0.1$ and $\nu_0 = 20/2\pi$, with regular sampling. PSD fits are shown in Figure~\ref{fig:periodic_PSD}. The GP MCMC posteriors are in pink, which the FFT/GLS posteriors are in green.}
    \label{fig:periodic_corner reg}
\end{figure*}

\begin{figure*}
    \centering
    \includegraphics[width=0.98\textwidth]{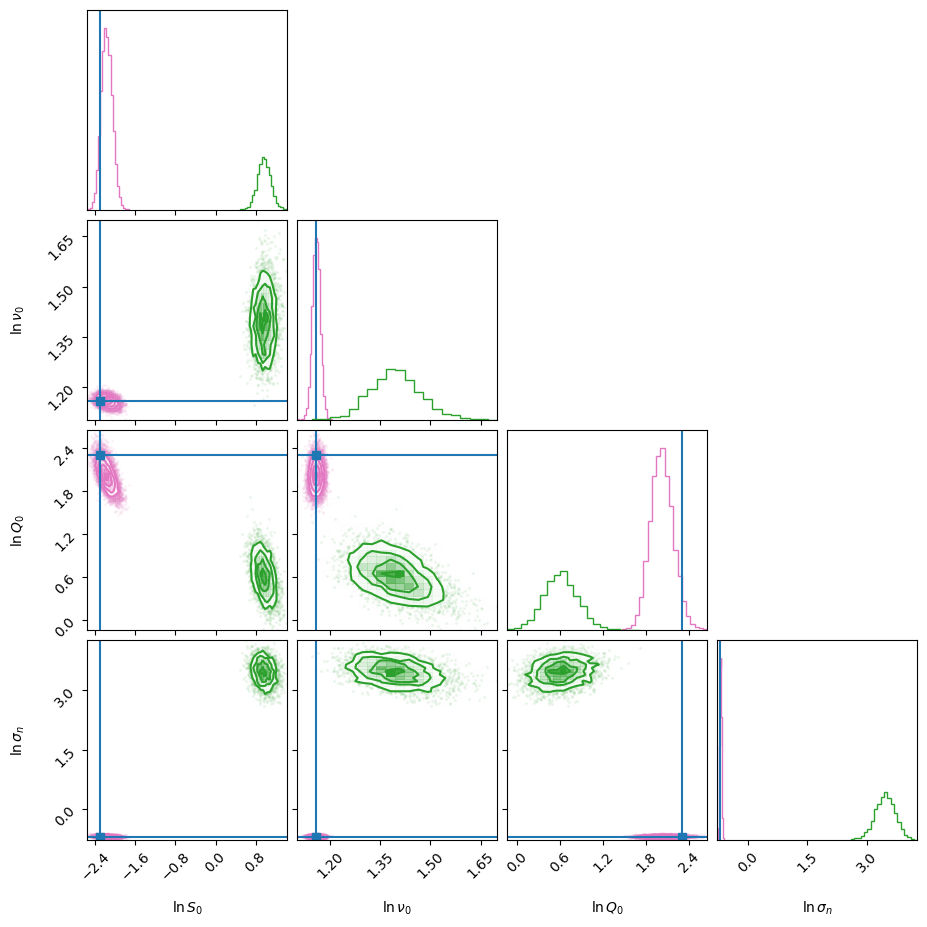}
    \caption{Corner plot for the case of a single, periodic component with $Q=10$ $S_0 = 0.1$ and $\nu_0 = 20/2\pi$, with irregular sampling. PSD fits are shown in Figure~\ref{fig:periodic_PSD}. The GP MCMC posteriors are in pink, which the FFT/GLS posteriors are in green.}
    \label{fig:periodic_corner itreg}
\end{figure*}

\begin{figure*}
    \centering
    \includegraphics[width=0.98\textwidth]{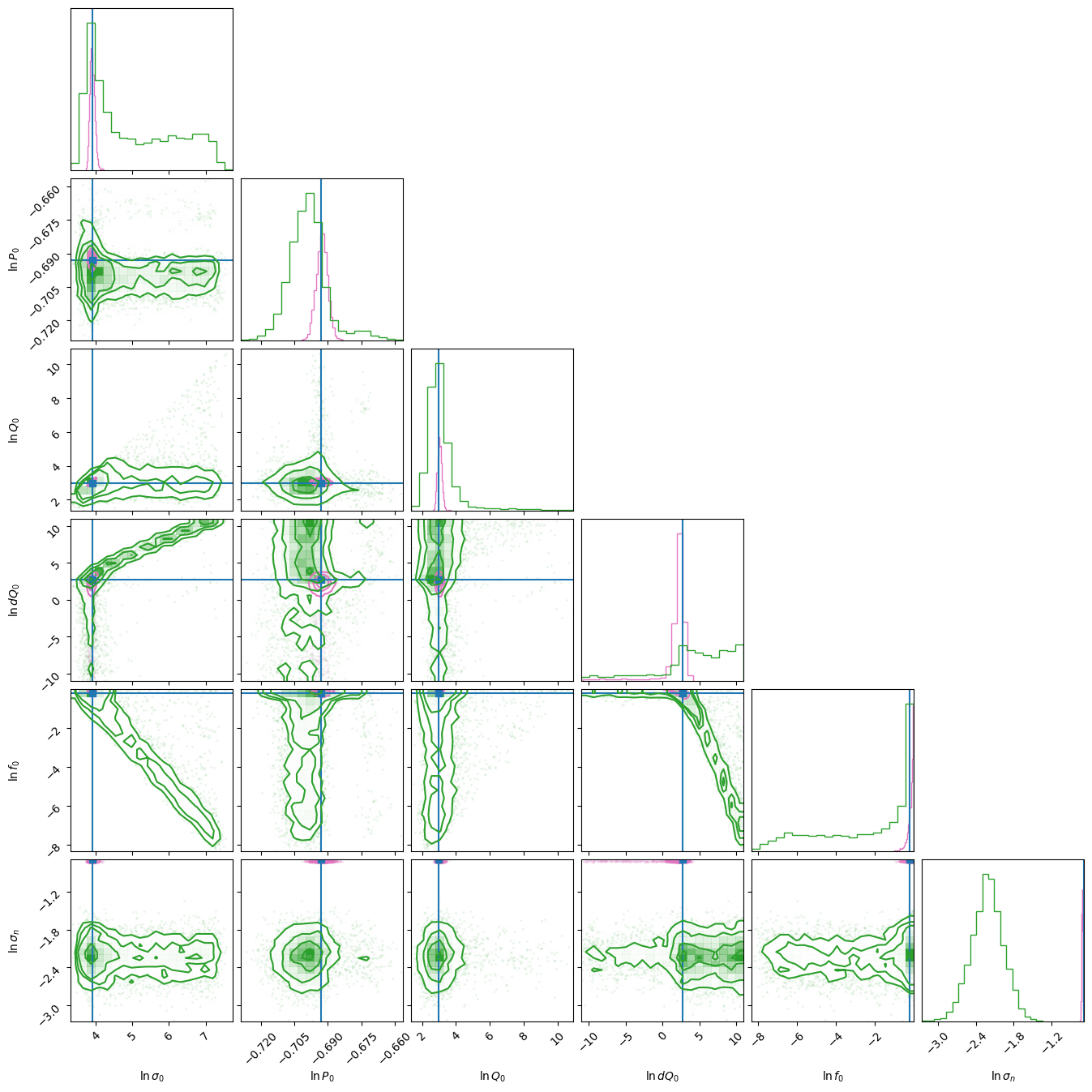}
    \caption{Corner plots for the \cel\ rotation term model in the regular time sampling case. The injected parameters are as follows : $\sigma =  50$, $P = 0.5$, $Q0 = 20$, $\delta Q = 15$, $f = 0.8$. The GP MCMC posteriors are in pink, which the FFT/GLS posteriors are in green.}
    \label{fig:rot corner reg}
\end{figure*}

\begin{figure*}
    \centering
    \includegraphics[width=0.98\textwidth]{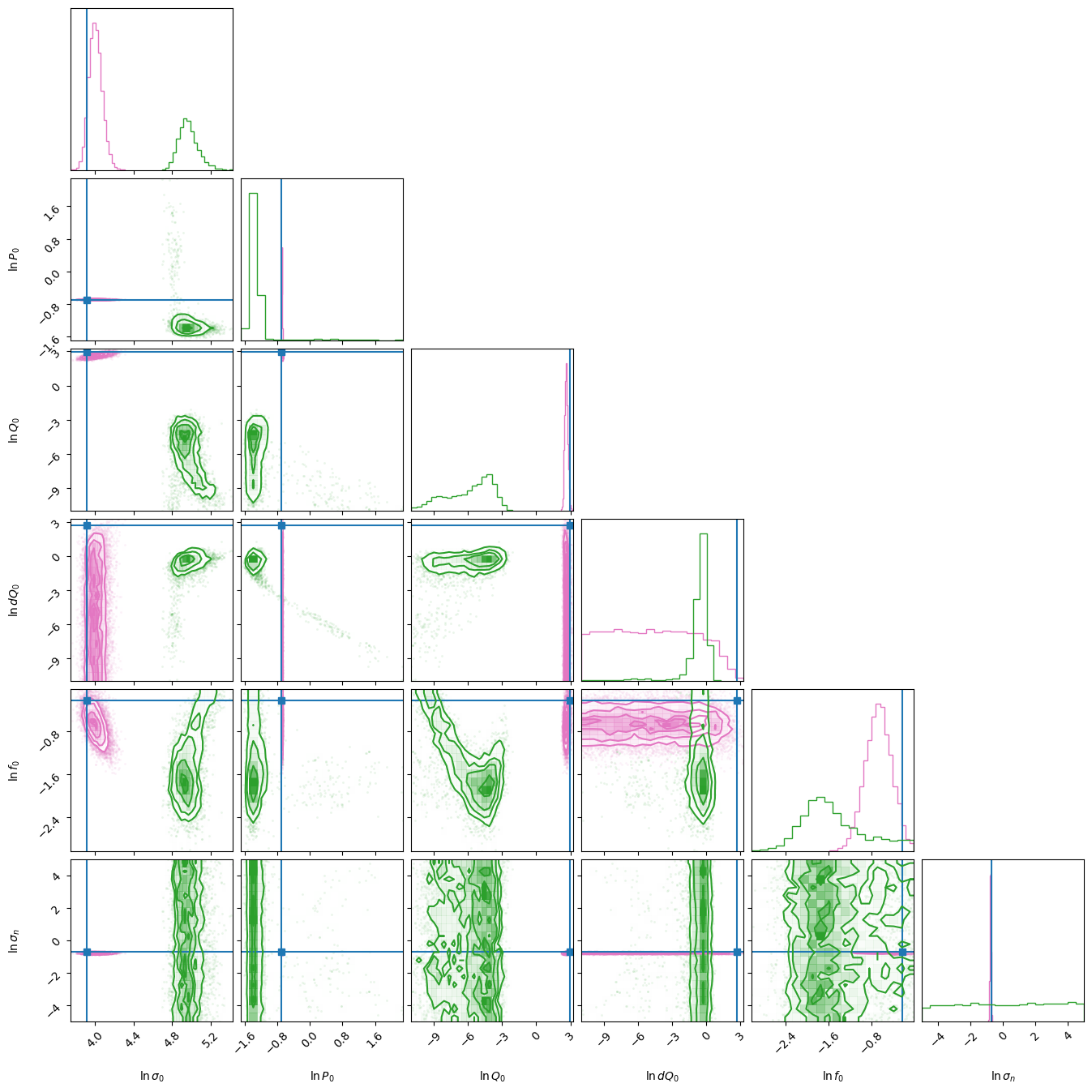}
    \caption{Corner plots for the \cel\ rotation term model in the irregular time sampling case. The injected parameters are as follows : $\sigma =  50$, $P = 0.5$, $Q0 = 20$, $\delta Q = 15$, $f = 0.8$. The GP MCMC posteriors are in pink, which the FFT/GLS posteriors are in green.}
    \label{fig:rot corner irreg}
\end{figure*}

\begin{figure*}
    \centering
    \includegraphics[width=0.98\textwidth]{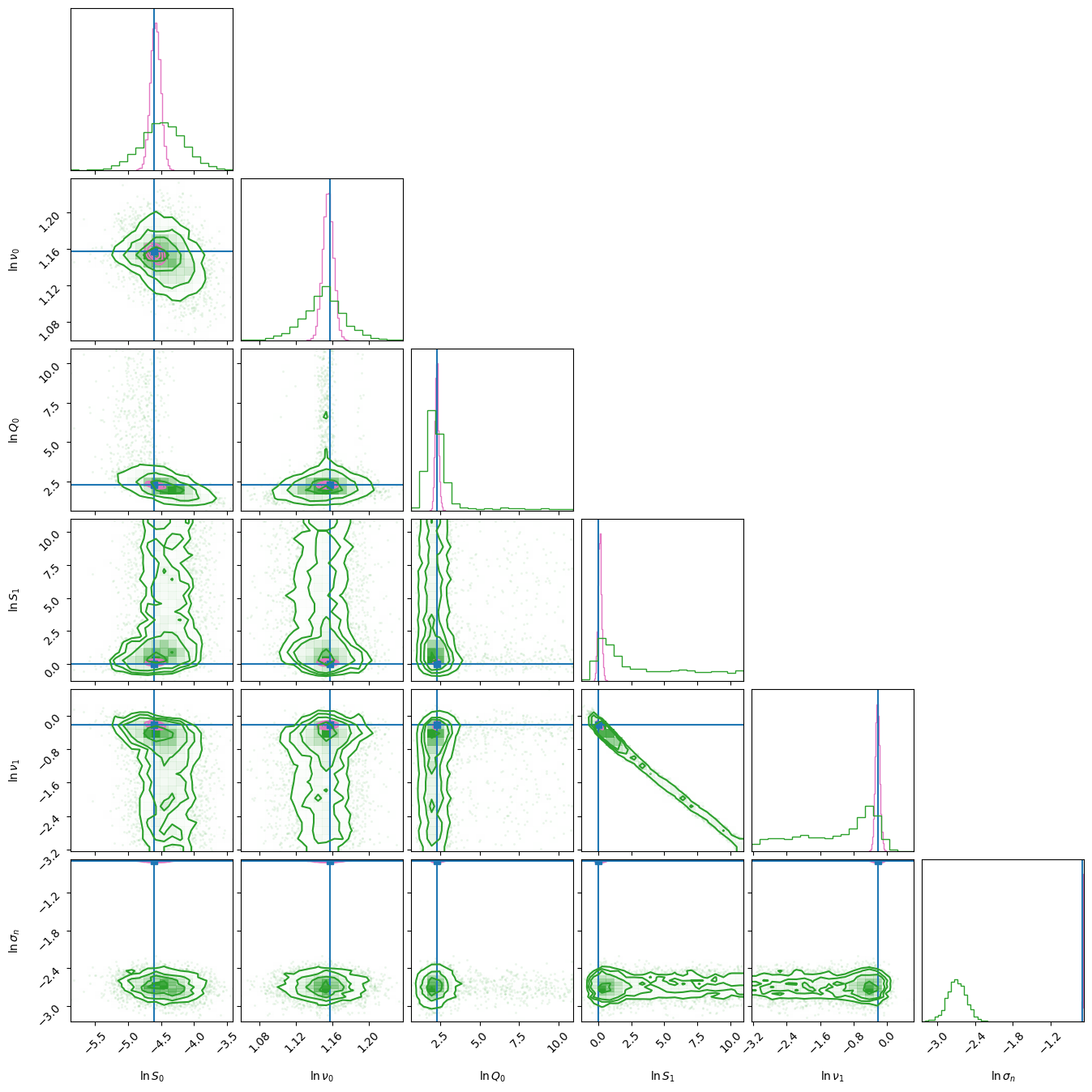}
    \caption{Corner plot for 2-component models with one periodic and one aperiodic component with regular time samping. The injected parameters are as follows : $S_1 =  0.01$, $\nu_1 = 20/2\pi$, $Q_1 = 10$, $S_2 = 1$, $\nu_2 = 5/2\pi$, $\sigma_n = 0.5$. The GP MCMC posteriors are in pink, which the FFT/GLS posteriors are in green.}
    \label{fig:aperiodic _ periodic_corner reg}
\end{figure*}

\begin{figure*}
    \centering
    \includegraphics[width=0.98\textwidth]{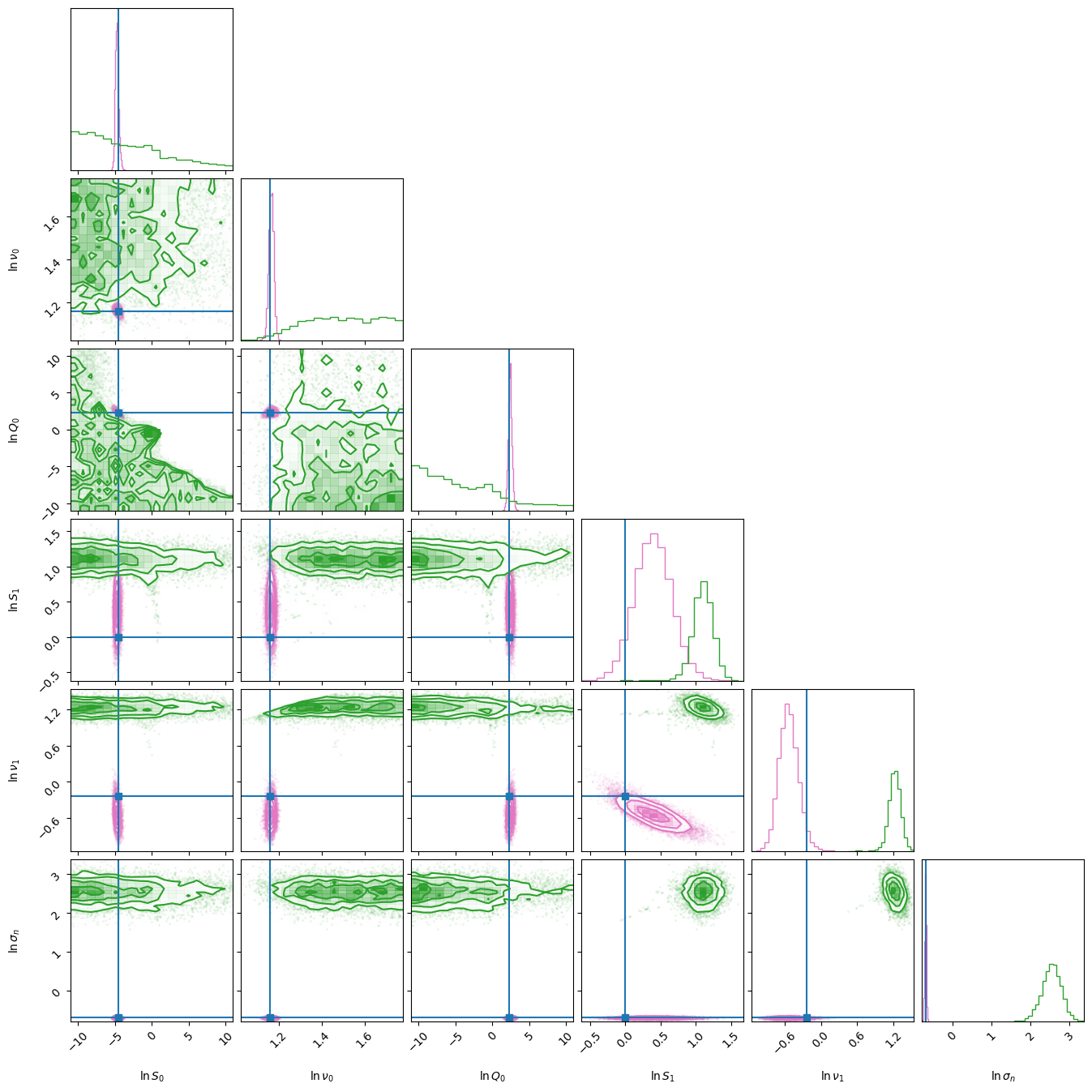}
    \caption{Corner plots for 2-component models with one periodic and one aperiodic component with irregular time sampling. The injected parameters are as follows : $S_1 =  0.01$, $\nu_1 = 20/2\pi$, $Q_1 = 10$, $S_2 = 1$, $\nu_2 = 5/2\pi$, $\sigma_n = 0.5$. The GP MCMC posteriors are in pink, which the FFT/GLS posteriors are in green.}
    \label{fig:aperiodic _ periodic_corner irreg}
\end{figure*}

\begin{figure*}
    \centering
    \includegraphics[width=0.98\textwidth]{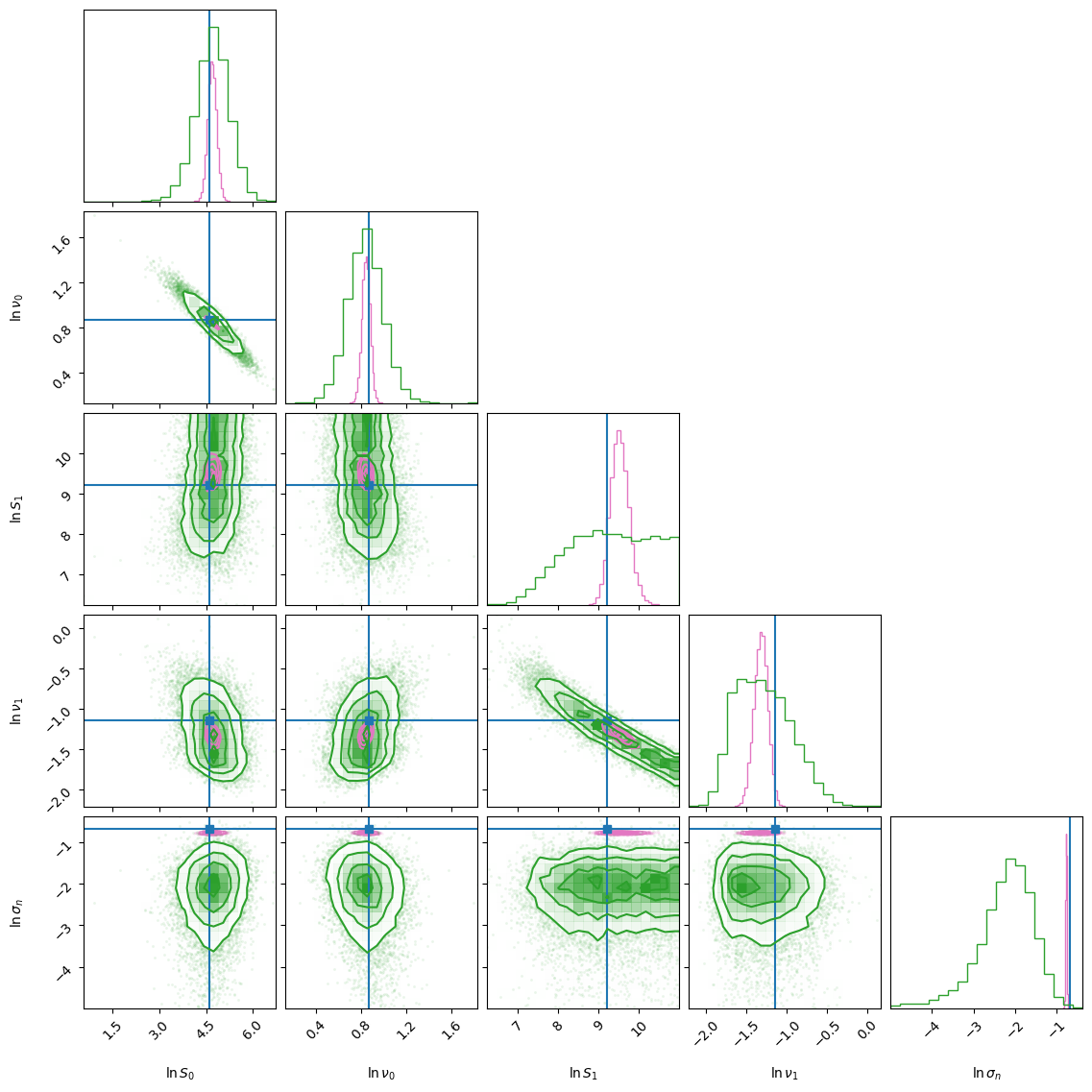}
    \caption{Corner plots for 2-component models with two aperiodic components with regular time sampling.  The injected parameters are as follows : $S_1 =  0.01$, $\nu_1 = 20/2\pi$, $Q_1 = 10$, $S_2 = 1$, $\nu_2 = 5/2\pi$, $\sigma_n = 0.5$. The GP MCMC posteriors are in pink, which the FFT/GLS posteriors are in green.}
    \label{fig:aperiodic _ aperiodic_corner reg}
\end{figure*}

\begin{figure*}
    \centering
    \includegraphics[width=0.98\textwidth]{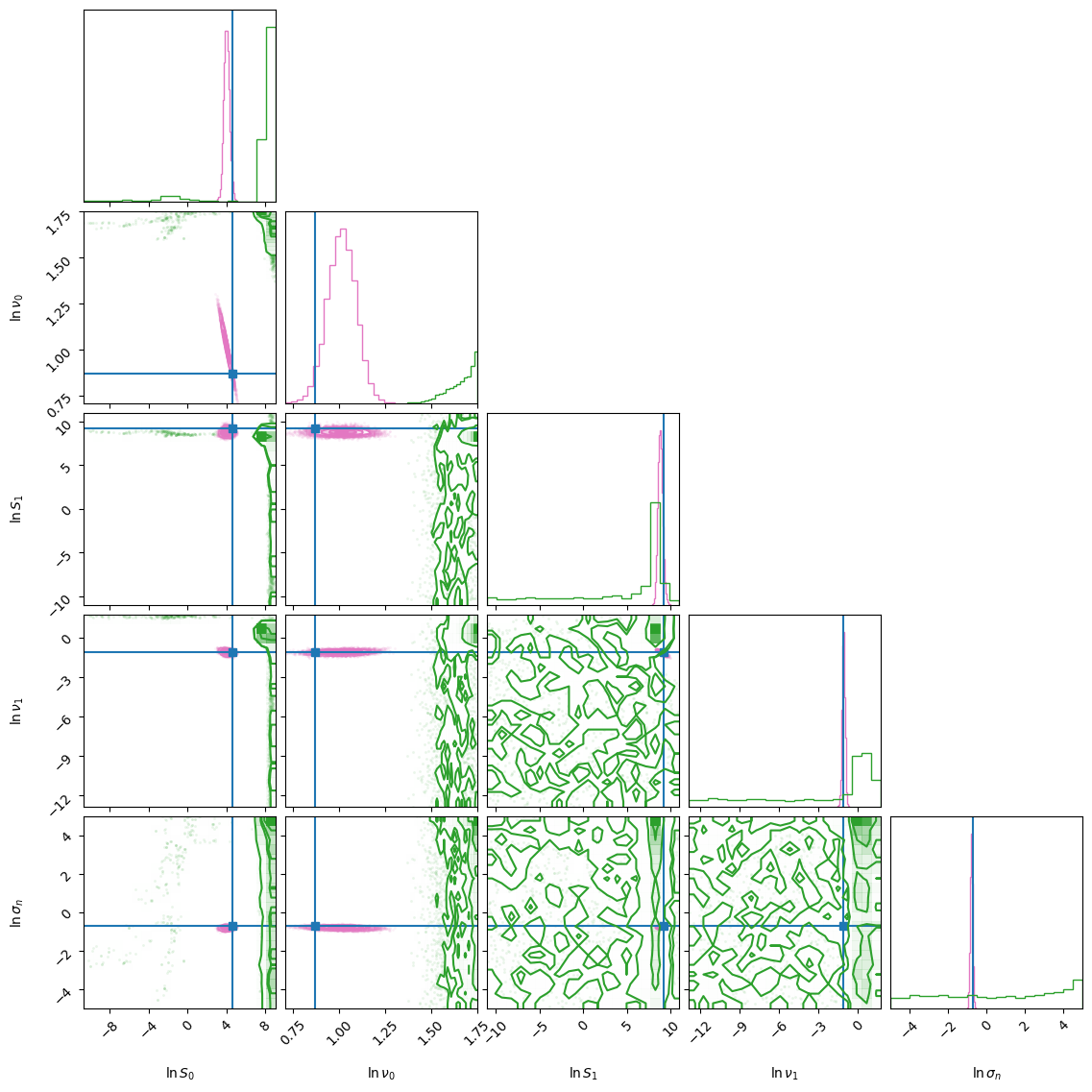}
    \caption{Corner plots for 2-component models with two aperiodic components with  irregular time sampling.  The injected parameters are as follows : $S_1 =  0.01$, $\nu_1 = 20/2\pi$, $Q_1 = 10$, $S_2 = 1$, $\nu_2 = 5/2\pi$, $\sigma_n = 0.5$. The GP MCMC posteriors are in pink, which the FFT/GLS posteriors are in green.}
    \label{fig:aperiodic _ aperiodic_corner irreg}
\end{figure*}

\bsp	
\label{lastpage}
\end{document}